\documentclass[final,5p]{elsarticle}

\usepackage{amsmath,amssymb,amsfonts}
\usepackage[]{algorithm2e}
\usepackage{graphicx}
\usepackage{textcomp}
\usepackage{soul}
\usepackage{color}
\usepackage{xcolor}
\usepackage{listings}
\usepackage[utf8]{inputenc}
\usepackage{mathtools}
\usepackage{booktabs}
\usepackage{multicol}
\usepackage{url}
\usepackage{comment}
\usepackage{makecell}
\usepackage{caption}

\def\uschema{\hbox{U-Schema}}

\def\uschema{\hbox{U-Schema}}
\lstdefinelanguage{pseudo} {
  basicstyle=\fontsize{8}{9}\selectfont\ttfamily,
  keywords={if,db,println,query},
  keywordstyle=\color{black}\bfseries,
  stringstyle=\color{black},
  morestring=[b]',
  morestring=[b]"
}

\lstdefinelanguage{js} {
  basicstyle=\fontsize{8}{9}\selectfont\ttfamily,
  keywords={typeof,new,true,false,catch,function,return,null,catch,switch,var,if,in,while,do,else,case,break, print, findOne,collection,db,const},
  keywordstyle=\color{blue}\bfseries,
  ndkeywords={class,export, boolean, throw, implements, import, this},
  ndkeywordstyle=\color{darkgray}\bfseries,
  identifierstyle=\color{black}, sensitive=false, comment=[l]{//},
  morecomment=[s]{/*}{*/}, commentstyle=\color{purple},
  stringstyle=\color{black},
  morestring=[b]',
  morestring=[b]"
}

\begin{document}
\begin{frontmatter}

\title{Towards the Automated Extraction and Refactoring of NoSQL Schemas from Application Code}

\author[UM]{Carlos J. Fernandez-Candel}
\ead{cjferna@um.es}
\author[UNamur]{Anthony Cleve}
\ead{anthony.cleve@unamur.be}
\author[UM]{Jesús J. Garcia-Molina}
\ead{jmolina@um.es}

\address[UM]{Faculty of Computer Science, University of Murcia, Murcia, Spain}
\address[UNamur]{Namur Digital Institute (NaDI), University of Namur, Namur, Belgium}

\begin{abstract}

Most NoSQL systems adopt a schema-on-read approach to promote flexibility and agility: the structure of the stored data is not constrained by predefined schemas. However, the absence of explicit schema declarations does not imply the absence of schemas themselves. In practice, schemas are implicit in both the application code and the stored data, and are essential for building tools such as data modelers, query optimizers, data migrators, or for performing database refactorings. As a result, NoSQL schema inference (also known as schema extraction or discovery)  has gained attention from the database community, with most approaches focusing on extracting schemas from data. In contrast, the source code analysis remains less explored for this purpose.

In this paper, we present a static code analysis strategy to extract logical schemas from NoSQL applications. Our solution is based on a model-driven reverse engineering process composed of a chain of platform-independent model transformations. The extracted schema conforms to the \uschema{} unified metamodel, which can represent both NoSQL and relational schemas. To support this process, we define a metamodel capable of representing the core elements of object-oriented languages. Application code is first injected into a code model, from which a control flow model is derived. This, in turn, enables the generation of a model representing both data access operations and the structure of stored data. From these models, the \uschema{} logical schema is inferred. Additionally, the extracted information can be used to identify refactoring opportunities. We illustrate this capability through the detection of join-like query patterns and the automated application of field duplication strategies to eliminate expensive joins. All stages of the process are described in detail, and the approach is validated through a round-trip experiment in which a  application using a MongoDB store is automatically generated from a predefined schema. The inferred schema is then compared to the original to assess the accuracy of the extraction process.

\end{abstract}


\end{frontmatter}

\section{Introduction}\label{sec:introduction}

Most NoSQL systems follow a “schema-on-read” approach, allowing data to be stored without a predefined schema. While this schemaless\footnote{Throughout this paper, we use the terms "schema-on-read" and "schemaless" interchangeably to refer to database systems that do not enforce a predefined schema at write time.} nature grants developers the flexibility to handle frequent changes in data structures --- common in modern applications --- it also introduces a key challenge: building database utilities such as schema visualization, code generation, and query optimization typically requires knowledge of the underlying data structure. As highlighted in~\cite{bacvanski-datav2015}, NoSQL database tools should incorporate reverse engineering strategies to extract the implicit schema from code or data. 

NoSQL systems are commonly classified into four categories based on their underlying data model: columnar, document, key–value, and graph as noted by~\cite{fowler-nosql2012}. In the first three models, aggregation tends to prevail over references between data, as discussed in~\cite{metamodel2021,fowler-nosql2012}. Specifically, key–value systems store data in associative arrays, document-oriented systems use JSON-like documents, and columnar systems organize data in tables where a column may contain aggregates of other columns. In contrast, graph databases are designed to store highly connected data, where the primary focus is on relationships between entities, and the aggregation of internal data is less relevant or sometimes unnecessary.

Both static code analysis of database applications and stored data analysis are traditional techniques to extract information from databases. In the context of NoSQL schema extraction, several  data analysis strategies have been proposed. Most of these approaches target a single data model. For instance, document stores such as MongoDB\footnote{MongoDB Database website: \url{https://www.mongodb.com/}} are addressed in~\cite{klettke-schema2015,sevilla-er2015,wang-schema2015}, graph databases like Neo4J\footnote{Neo4j Database website: \url{https://www.neo4j.com}}, are considered in~\cite{comyn-wattiau2017}, and columnar databases such as HBase\footnote{HBase Database website: \url{https://hbase.apache.org/}} are covered in~\cite{frozzaDM21}.

More recently, Carlos J. Fernández-Candel et al. presented a strategy based on the \uschema{} unified metamodel, which can represent schemas for both relational and NoSQL databases~\citep{metamodel2021}. This metamodel differs from the schema representations used in previous proposals in three main aspects: (i) it supports both aggregation and reference relationships between entity types; (ii) it explicitly models relationship types, as required in graph schemas or many-to-many tables in relational schemas; and (iii) it allows entity and relationship types to have multiple structural variations, which is essential in schemaless environments where a single fixed structure per type is not enforced.

As far as we know, no approaches applying code analysis to infer NoSQL complete schemas have been published. However, static code analysis of MongoDB applications has been used for related purposes, such as discovering entity evolution and detecting database access operations~\citep{meurice2017,cherryBGMNC22}. 
\cite{meurice2017} presented a strategy aimed at tracking how the schema of a specific entity type evolves throughout an application's lifecycle. Their approach analyzes different versions of Java applications to identify structural changes over time —i.e., entity versions. However, it does not extract a complete database schema, as relationships between entities are not identified.
Boris Cherry et al~\citep{cherryBGMNC22}. proposed a method to detect MongoDB access operations in JavaScript applications. Their work focuses on locating access points—such as queries, insertions, updates, and deletions—across diverse codebases.

While data reverse engineering was used in~\cite{metamodel2021} to infer NoSQL logical schemas as \uschema{} models, in this work we present a static code analysis strategy with the same objective. A relevant distinction should be made between both approaches. Data-driven strategies are well suited to detecting structural variability in schema types, whereas identifying such variability through code analysis is considerably more complex. Therefore, our strategy is complementary to that presented in~\cite{metamodel2021}, as both approaches extract NoSQL logical schemas as \uschema{} models. In our case, schema type versions can be detected in a manner similar to the approach of ~\cite{meurice2017}, by analyzing multiple versions of the application over time.

Our reverse engineering solution is implemented as a three-step model transformation chain. First, the source code is injected into a model that captures the structure of programming language statements. From this code model, a control flow model is derived, preserving references to the original code elements. This control flow model is then traversed to generate a model that captures CRUD operations and the physical structure of the data, referred to as the Database Operations and Structure (DOS) model. Finally, the \uschema{} logical schema is obtained from the physical schema embedded within the DOS model. 

Beyond schema extraction, we also explored the use of the DOS model to support the automation of database refactorings in NoSQL environments. In particular, we focused on identifying join queries to provide database practitioners~\footnote{In the context of this work, we use the term ''database practitioners`` to collectively refer to database administrators and developers involved in schema design or refactoring tasks.} with actionable information for deciding whether a specific join can be eliminated by duplicating fields from the referenced entity into the referencing one. This operation corresponds to a well-known denormalization technique used in relational databases to optimize query performance—for instance, by storing detail records together with master data. A similar strategy is considered a best practice in document databases such as MongoDB\footnote{MongoDB Best Practices: \url{https://www.mongodb.com/basics/best-practices}}, especially when applications perform queries involving references (i.e., joins) between two document collections. Once a candidate join is selected for removal, the corresponding schema change is automatically applied: the schema, data, and source code are updated accordingly.

Each step of the reverse engineering process was tested by verifying the correctness of the generated models. To this end, several validation strategies were applied, including:~(i)~rewriting the source code represented in the models,~(ii)~generating textual representations to facilitate the identification of database-related statements; and~(iii)~visualizing the control flow and the extracted database schema using tables and graphs. The overall process was validated through a round-trip experiment, in which JavaScript code accessing a MongoDB store  was analyzed. This code is part of an automatically generated Node.js application. The extracted logical schema was compared both with the one originally designed, as well as with the one schema inferred directly from the stored data.

Our work contributes to the state of the art in the following ways:

\begin{itemize}% [topsep=2pt, itemsep=2pt, parsep=0pt]
	\item To the best of our knowledge, this is the first code analysis approach capable of extracting NoSQL logical schemas, where entity types are connected through both aggregation and reference relationships. Additionally, the approach supports relationship types. %and structural variations. 
    By representing schemas using a generic metamodel, our method becomes applicable to multi-model data-base tools, enabling broader use in diverse database environments.
	
	\item The extracted information in the devised reverse engineering process also enables the automation of data-base refactorings. In this paper, we illustrate this capability with a strategy to duplicate fields in order to eliminate the need for expensive joins, thus improving query performance.
	
	\item The proposed reverse engineering process leverages metamodel-based abstractions~\citep{brambilla2012} to represent the involved information: source code, control flow, data-base operations, and data structures. The metamodels were designed to ensure platform independence and reusability. In particular, the code metamodel captures common elements of object-oriented languages and can be extended with language-specific constructs.
    
    \item A controlled round-trip validation strategy was conducted. A large language model (LLM) was used to generate a Node.js application from a database schema. This validation allowed for precise comparison between the extracted schema and the original design, and enabled the analysis of join patterns and the verification of refactoring outcomes in a realistic yet reproducible setting.
    
\end{itemize}

The full implementation of our approach, which includes all metamodels, model-to-model transformations, and code analysis algorithms used for generating intermediate models, extracting schemas, and automating refactorings, is publicly available on GitHub~\footnote{U-Schema: Code analysis Inference and Refactoring - ModelUM Repository: \url{https://github.com/modelum/uschema-code-analysis}}.

This paper is structured as follows. Section~\ref{sec:overview} provides an overview of the proposed approach, which is described in detail across Sections~\ref{sec:representingcode} to~\ref{sec:generatingplans}. Section~\ref{sec:representingcode} presents the internal representation of source code using the Code and Control Flow metamodels. Section~\ref{sec:generatingDOS} describes the construction of the DOS model from these representations. Section~\ref{sec:obtainingschemas} explains how \uschema{} models are derived from DOS models, while Section~\ref{sec:generatingplans} focuses on the identification of join queries and the application of field duplication refactorings. All metamodels and algorithms are described in detail, including the testing strategies applied to each of them. Section~\ref{sec:validation} presents the overall validation of the complete approach. Finally, Section~\ref{sec:relatedWork} reviews related work, and Section~\ref{sec:conclusions} concludes the paper and outlines future research directions.

\section{Overview of the approach\label{sec:overview}}

This section outlines the approach presented in this paper. A running example involving an operation on a document store  is used to illustrate the kind of information that can be discovered through code analysis to extract the logical schema and automate database refactoring.

In document stores, semi-structured objects are stored in collections, and each object has a JSON-like structure consisting of a set of name–value pairs. The value of a field can be a primitive type, an array of objects, or an embedded object. For example, Figure~\ref{fig:runningExampleJsons} shows sample \textit{user} and \textit{movie} objects in a document store used by a streaming service to manage subscribed users, including personal information and a list of watched movies.

User objects contain a \texttt{watchedMovies} field, which holds an array of embedded objects. Each embedded object includes two fields: \texttt{movie\_id}, which stores the identifier of the watched movie (i.e., a reference); and \texttt{stars}, which contains the user’s rating of that movie.

\begin{figure}[!htb]  
\centering
%\vspace*{-1em}
\begin{lstlisting}[
	language=js,
	frame=single
]
// User Collection
{
  "_id": 101,
  "name": "Brian",
  "surname": "Caldwell",
  "email": "brian_caldwell@gmail.com",
  "watchedMovies": [
    {
      "stars": 7,
      "movie_id": 202
    },
    {
      "stars": 10,
      "movie_id": 303
    }
  ]
}

// Movie Collection
{
  "_id": 202,
  "title": "The Matrix"
  "director": "The Wachowskis"
}
{
  "_id": 303,
  "title": "The Godfather",
  "director": "Francis Ford Coppola"
}
\end{lstlisting}
\caption{Users and Movies objects in the ``streaming service'' document store.
	\label{fig:runningExampleJsons}}
\end{figure}

Listing~\ref{code:originalCodeRunningExample} presents a pseudo-code example of an operation referred to as “First-Watched-Movie” (FWM), which consists of three statements. First, a query selects a user by name. Then, a join query retrieves the first movie watched by that user. Finally, if the user rated that movie with a number of stars greater than or equal to $5$, the following information is printed to the console: the user's full name, as well as the title and star rating of the retrieved movie. 

\begin{figure}[!ht]
\begin{lstlisting}[
    language=pseudo,
    numbers=none,
    frame=single,
    label={code:originalCodeRunningExample},
    caption={Pseudo-code of the FWM database operation.},
    captionpos=b,
]
1. user = db("Users").query(name == "Brian")
2. movie = db("Movies").query(
    _id == user.watchedMovies[0].movie_id)
3. if (user.watchedMovies[0].stars >= 5)
  println user.name+user.surname+user.email
  println "Last watched movie:"
  println movie.title+user.watchedMovies[0].stars

\end{lstlisting}
\end{figure}

To facilitate code analysis, a more abstract representation is derived from the AST produced by a parser. We define a set of interconnected models: one that captures the control flow of the program, and another that represents its abstract syntax. Nodes and edges in the Control Flow model are linked to elements in the Code model, enabling the analysis of execution paths while preserving structural context. The process begins by injecting the source code into a Code model, from which the Control Flow model is constructed, as described in Section~\ref{sec:representingcode}.

Extracting the logical schema and automating database refactorings requires identifying both the physical structure of the data and the CRUD operations performed on it. This information is captured in the DOS model, which is obtained by traversing the Control Flow model, as explained in Section~\ref{sec:generatingDOS}. Specifically, the traversal must visit relevant statements to identify the following elements:

\begin{itemize}%[topsep=2pt, itemsep=2pt, parsep=0pt]
	\item  \textit{Data containers} (e.g., collections in document stores 
	or tables in relational or columnar databases). They can be identified 
	through the analysis of CRUD operations, e.g., by inspecting the argument 
        of \texttt{db()} calls in the queries of the FWM example.

         \item \textit{Structure of objects} stored in a particular container, 
         defined as a set of properties represented by name –type pairs. 
         Properties are inferred from object fields, which can be detected in expressions 
         that use dot notation to access a variable’s attributes, 
         such as \texttt{user.name} or \texttt{movie.title}.
	
	\item \textit{Variables holding database objects}. These are discovered by analyzing 
	expressions such as method invocations, assignments, and arguments. For example, the
	variables \texttt{user} and \texttt{movie}  are detected in the assignments found in
	statements 1 and 2 of the FWM example.

        \item \textit{Types of properties}. Properties can be of primitive type, 
        collection (e.g., array or list), aggregate, or reference. Primitive types 
        are inferred from expressions such as assignments or conditions 
        (e.g., \texttt{name == "Brian"}). An array type is detected when elements 
        are accessed using index notation. For example, analyzing the expression 
        \texttt{user.watchedMovies [0].movie\_id} reveals that the \texttt{watchedMovies} 
        property is an array. Furthermore, it indicates that array elements are objects 
        containing the \texttt{movie\_id} property. In such cases, the type of 
        \texttt{watchedMovies} is inferred as an array of an aggregate type, 
        typically named after the field in singular form. We use the term 
        \textit{non-root entities} to refer to these aggregate types, 
        distinguishing them from root entities corresponding to data containers 
        (e.g., \texttt{user} and \texttt{movie}).
	
	\item \textit{CRUD operations}. These are detected by identifying calls to database 
	API functions. In the FWM example, assuming the \texttt{query()} method
	issues a read operation on the store specified by the \texttt{db()} call, 
        two read operations would be identified in statements 1 and 2.

        \item \textit{Reference and join queries}. A query is identified as a join 
        when its condition includes an equality check between the identifier field 
        of one object and a field of a previously retrieved object. In such cases, 
        the latter field is considered a reference, and the corresponding property 
        is assigned a reference type. In the FWM example, the query on movies (statement 2) 
        qualifies as a join query, and the type of the \texttt{movie\_id} property
        is inferred as “Reference to Movie.”
   
\end{itemize}

\begin{figure}[!ht]
  \centering
  \includegraphics[width=\linewidth]{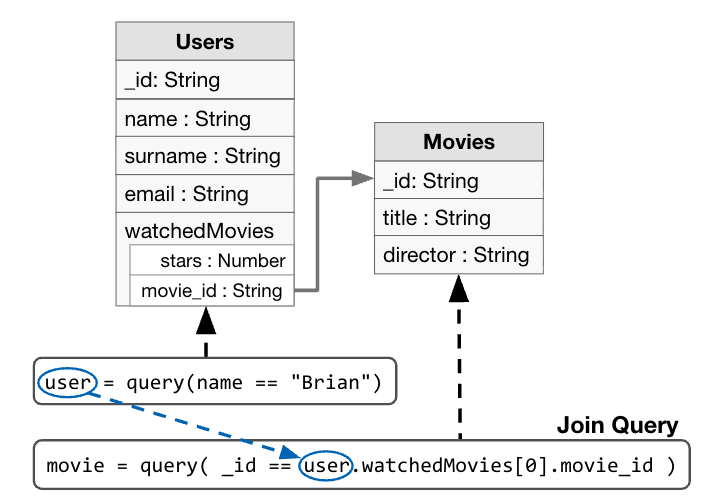}
	\caption{Entities and Queries extracted for the FWM example.}
	\label{fig:schemaRunningExample}
\end{figure}

Figure~\ref{fig:schemaRunningExample} illustrates the information contained in the DOS model for the pseudo-code of the FWM example. The physical data structure is shown at the top of the figure, while the queries appear at the bottom.
Note that the types of the fields \texttt{surname} and \texttt{email} in \textit{User}, and \texttt{title} and \texttt{director} in \textit{Movie}, cannot be determined directly from the pseudo-code. Since these fields only occur in print statements, the default type \texttt{String} is assigned to them. In our approach, the physical data schema is then transformed into a logical schema represented using the \uschema{} generic metamodel, as described in Section~\ref{sec:obtainingschemas}.

The DOS model can also be used to detect candidate database refactorings. In this paper, we illustrate this capability by focusing on the join query removal refactoring. A join query involves four elements: a source container, a target container, a query on the source container, and the join condition used to select the object from the target container.
Removing a join query is possible if the relevant properties of the target entity are copied (i.e., duplicated) into the source entity. In this way, the query on the source container becomes sufficient to retrieve all the required information. However, not all properties of the target entity need to be copied, only those that are actually accessed in the code following the join query.
In our example, the \texttt{title} field should be copied into the \textit{WatchedMovie} objects, but not the \texttt{director} field, since the statement following the query accesses \texttt{movie.title} but not \texttt{movie.director}.

Therefore, we propose a code analysis approach to identify the data that should be duplicated for each join query. To  this end, the list of join queries is iterated, and for each one, subsequent statements are inspected to determine which fields need to be copied. This analysis provides database practitioners with with actionable insights to support decisions on which join queries can be removed. These insights includes: the source and target containers, the join query itself, the original and modified versions of the query on the source container, the number of lines in which the retrieved data is used, and other queries involving the same containers —allowing database practitioners to assess how frequently the duplicated data is updated. All the information collected during this analysis is referred to as a ``join query removal plan''.

Like any schema change operation, the data duplication involved in a join query removal refactoring requires updating the schema, the database, and the application code. We have automated this process as follows:
(i) the logical schema is modified by adding the duplicated attributes from the referenced entity to the referencing entity;
(ii) the database is updated by inserting the duplicated fields to all referencing objects; and
(iii) the code is rewritten to remove the join query and to replace all references to the duplicated fields with direct accesses to the updated object.

In the case of the FWM script:
(i) the \textit{title} attribute is added to the \textit{WatchedMovie} entity type in the schema, through an operation on the \uschema{} model;
(ii) the \texttt{watchedMovies} array of each \textit{user} object is updated so that each embedded object includes the  \texttt{title} field from the referenced movie; and
(iii) the FWM script is rewritten as shown in Listing~\ref{code:resultCodeRunningExample}.

\begin{figure}[!ht]
\begin{lstlisting}[
    language=pseudo,
    numbers=none,
    frame=single,
    label={code:resultCodeRunningExample},
    caption={Pseudo-code updated when join query is removed.},
    captionpos=b,
]
1 user = db("Users").query(name == "Brian")
2 if (user.watchedMovies[0].stars >= 5)
3  println user.name + user.surname + user.email
4  println "Last watched movie": 
5  println user.watchedMovies[0].movie_title
           + user.watchedMovies[0].stars
\end{lstlisting}
\end{figure}

\begin{figure}[!ht]
  \centering
  \includegraphics[width=\linewidth]{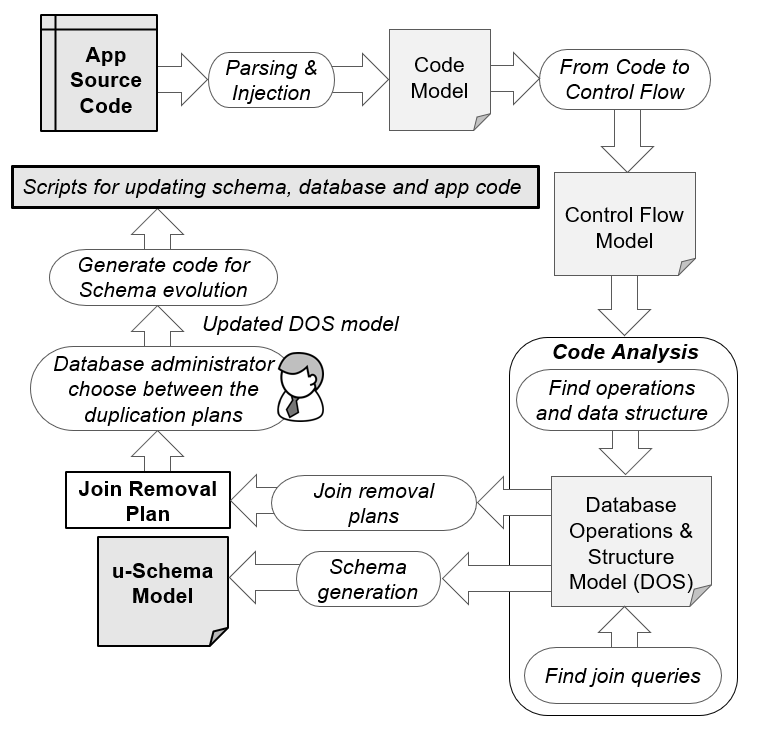}
  \caption{Overview of the \uschema{} extraction and join query removal approach.\label{fig:overviewImplementation}}
\end{figure}

Figure~\ref{fig:overviewImplementation} illustrates the sequence of stages in the strategy outlined above. The source code is first injected into a Code model, from which a Control Flow model is derived. This Control Flow model is then analyzed to generate the DOS model, which serves as input to two subsequent processes: one that transforms the physical schema into a logical schema, and another that generates join query removal plans. These plans are presented to database practitioners, who select which ones to apply. For each selected plan, the schema, database, and application code are updated accordingly.

Listing~\ref{code:running-example} presents the FWM pseudo-code expressed in JavaScript, which will be used as a running example in the following sections. We assume that MongoDB is the underlying document store.
lines~1~to~7 initialize the \texttt{client} variable, which holds the client-side connection to a MongoDB database. 
Line 9 marks the beginning of the code corresponding to the pseudo-code shown in Listing~\ref{code:originalCodeRunningExample}. It is implemented as a \texttt{findOne()} query on the \texttt{Users} collection, which takes two arguments: a query condition and a callback function (i.e., a lambda expression) whose parameter is the object returned by the query. This callback includes a nested \texttt{findOne()} query on the \texttt{Movies} collection. In this inner query, the first argument defines the join condition, while the second is another callback containing an \textbf{if-then} statement. The body of this statement consists of three consecutive \texttt{console.log} statements. As a result, the example involves a nesting of three code blocks.

\begin{figure}[!ht]
\begin{lstlisting}[
language=js,
frame=single,
label={code:running-example},
caption={JavaScript code for the pseudo-code in Listing~\ref{code:originalCodeRunningExample} (data stored in MongoDB).},
captionpos=b
]
1 const MongoDB = require('mongodb').MongoClient;
2 
3 const url = 'mongodb://modelum.es/db:27017';
4 const dbName = 'streamingservice';
5
6 const client = new MongoDB(url);
7 client.connect(err => {
8
9 client.db(dbName).collection('users').findOne(
10 { name: 'Brian' }, (err, user) => {
11 
12 client.db(dbName).collection('movies').findOne( 
13  { _id: user.watchedMovies[0].movie_id }, 
14  (err, movie) => {
15    if (user.watchedMovies[0].stars >= 5) {
16     console.log(user.name +' '+ user.surname);
17     console.log(user.email + 
18                 ' Last watched movie:');
19     console.log(movie.title + ' ' + 
29                  user.watchedMovies[0].stars);
30    }
31   });
32  });
33 });
\end{lstlisting}
\end{figure}

\section{Obtaining an abstract representation of the code\label{sec:representingcode}}

Following the principles of Model-Driven Engineering (MDE)~\citep{fowler-dsl2010,greenfield-2004,volter-book}, we represent the source code to be analyzed through two complementary metamodels, which are described in this section. The Code metamodel captures the structural elements of the application—such as classes, methods, and statements—while the Control Flow metamodel represents the execution order of these statements. These representations provide the foundation for subsequent analysis stages, including the identification of database operations and the extraction of the schema. 

\begin{figure*}[!htb]
  \centering
  \includegraphics[width=\textwidth]{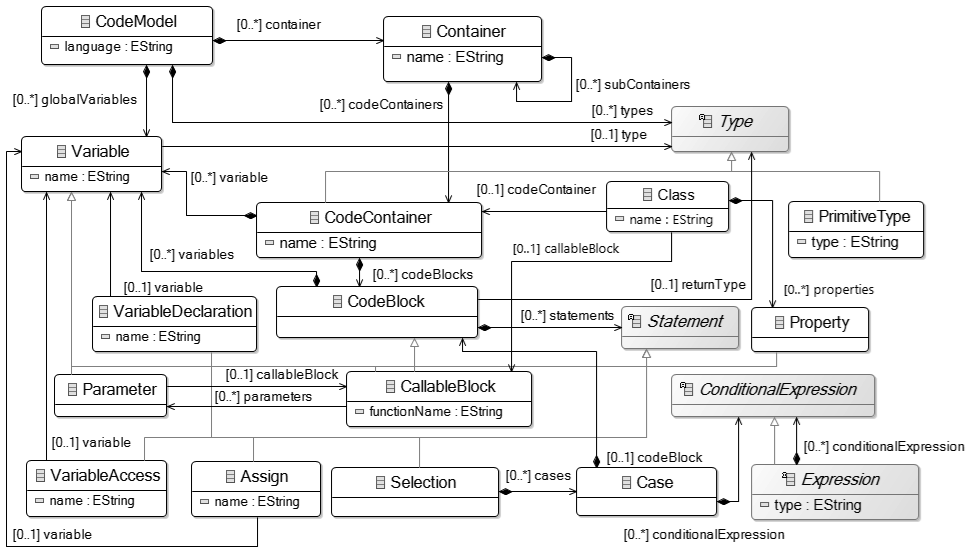}
  \caption{Excerpt of the main elements of the Code Metamodel.\label{fig:codeMM}}
\end{figure*}

\subsection{From source code to Code models \label{injectingcode}}

Building on our prior experience with the Java metamodel of the MoDisco framework and the Code metamodel defined in the OMG’s Knowledge Discovery Metamodel (KDM) specification~\citep{carlos-idioms19}, the Code metamodel was designed to provide an abstract and uniform representation of programs, independent of any specific programming language. The former helped us identify and formalize the main object-oriented constructs, while the latter inspired the design of a language-independent representation.

The metamodel captures the the main concepts and relationships that are common to object-oriented languages —such as \textit{Statement}, \textit{MethodInvocation}, and \textit{Class}— providing a lightweight and language-independent representation well suited for schema inference across different programming languages. An excerpt of the metamodel is shown in Figure~\ref{fig:codeMM}. The metamodel was not tailored to any specific language. Instead, language-specific elements could be defined in separate extension metamodels, allowing for a modular architecture easily adaptable to different programming languages. This design also ensures that the most common statements are centralized in the core \textit{Code} model, promoting consistency and reuse across language variants.

Next, we describe the \textit{Code} metamodel in sufficient detail to explain how \textit{Control Flow} models are derived. A \textit{Code model} represents an executable unit, such as a JavaScript script or program. It aggregates \textit{Container}s, global variable declarations (\textit{VariableDec}), as well as \textit{Type}s. These can be either primitive types or classes. Each \textit{Variable} has an associated \textit{Type}. This type is either taken from the variable declaration or locally inferred from expressions such as variable initializations or literal comparisons. The inference of more complex typing information is part of the analysis process and remains outside the scope of this work.
A \textit{Container} represents a structural unit that holds scripts or classes, such as packages, namespaces, folders, or files. Containers can be nested and may aggregate \textit{CodeContainer}s, which in turn group code blocks, class declarations, and variable declarations. 
A \textit{CodeBlock} contains an ordered list of \textit{Statement}s, such as conditionals or loops, and also declares local variables. Figure~\ref{fig:codeMM} shows only the types of statements that appear in the running example: conditional selection, method calls, variable accesses, and object creation expressions.
A special kind of code block is the \textit{CallableBlock}, which represents executable blocks invoked by the program, such as methods, functions, constructors, and lambda expressions.

\begin{figure}[!htb]
  \centering
  \includegraphics[width=\linewidth]{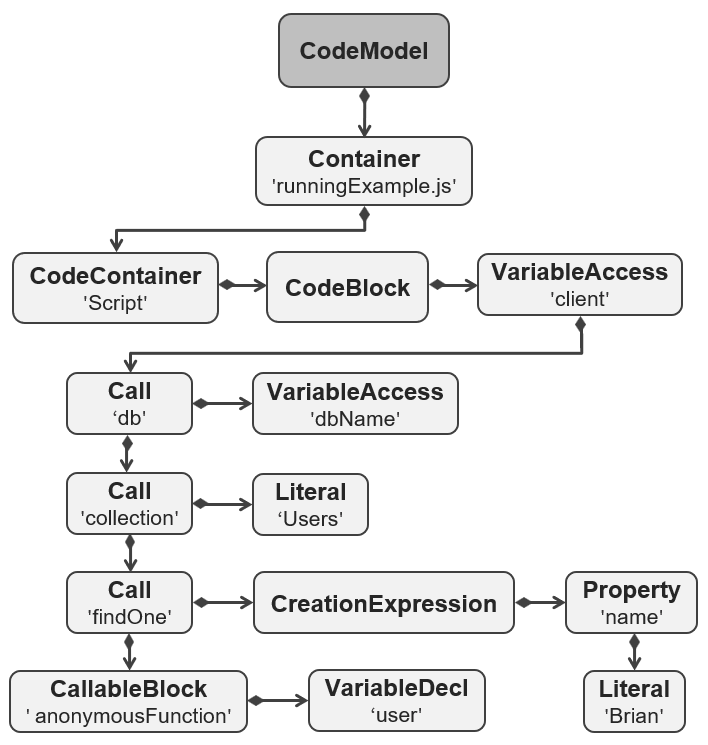}
  \caption{ Excerpt of the \textit{Code} model extracted for the running example (starting from line 9). \label{fig:codeModelRunningExample}}
\end{figure}

Figure~\ref{fig:codeModelRunningExample} shows an excerpt of the \textit{Code} model injected from the JavaScript code of the running example. This figure omits the portion of the model starting from the \textit{CallableBlock} element, which represents the lambda expression passed as the second argument of the \texttt{findOne()} query.
The injected model has a \textit{CodeModel} as its root element, which aggregates the container created for the script file \texttt{runningExample.js}. This container stores the absolute path of the file. For this script, a \textit{CodeContainer} of type ``script'' is created. This container aggregates a single \textit{CodeBlock} that corresponds to the script’s body, and it represents the dot notation expression that includes the outer \texttt{findOne()} call.
In this expression, the \texttt{client} variable is accessed to invoke the \texttt{db} method with the \texttt{dbname} argument (another \textit{VariableAccess} element). That call is then chained to the \texttt{collection} method with the string literal \texttt{'Users'} as an argument, and finally to the \texttt{findOne()} call. This last call takes two arguments: a lambda expression represented by an anonymous \textit{CallableBlock}, and an object creation expression that includes the property \texttt{'name'} and the literal value \texttt{'Brian'}.
The complete \textit{Code} model includes two \textit{Class} elements that represent the \texttt{user} and \texttt{movie} objects. Each class defines a set of properties: \texttt{movie} includes \texttt{\_id}, \texttt{title}, and \texttt{director}; while \texttt{user} includes \texttt{name}, \texttt{surname}, \texttt{email}, and \texttt{watchedMovies}.

It is worth noting that the injector responsible for obtaining Code models from the AST was manually implemented, rather than automatically generated using a language workbench. This decision was motivated by the complexity and size of the Code metamodel.

\paragraph{Testing}
To validate this stage, we applied the testing strategy defined in~\citep{carlos-idioms19}. We executed simple tests on minimal code snippets representing specific constructs (e.g., loops), automatically regenerating code from the corresponding Code model and comparing it to the original using a text-diff tool. This iterative process helped verify the correctness of the model injection logic for each construct. 

%%%%%%%%%%%%%%%%%%%%%%%%%%%%%%%%%%%%%%%%%%%%%%%%%%%%%%%%%%%%%%%%%%%%%%%%%%%%%%
%%%%%%%%%%%%%%%%%%%%%%%%%%%%%%%%%%%%%%%%%%%%%%%%%%%%%%%%%%%%%%%%%%%%%%%%%%%%%%

\subsection{Representing the control flow\label{control-flow}}

\begin{figure*}[!htb]
  \centering
  \includegraphics[width=\textwidth]{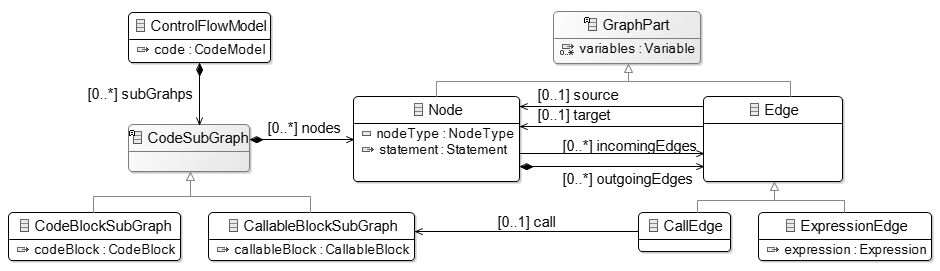}
  \caption{Control Flow Metamodel.\label{fig:graphMM}}
\end{figure*}

Code analysis typically requires not only a representation of the syntax tree, but also knowledge of the control flow graph. To represent control flow in our approach, we have defined the metamodel shown in Figure~\ref{fig:graphMM}. This metamodel defines the abstractions required to represent control-flow information obtained through the Ullman algorithm \cite{aho86}, following the representation strategy we previously applied to PL/SQL triggers \cite{oscar2011eventhandlers}. A \textit{Control Flow} model is derived from a \textit{Code} model, with its nodes and edges referencing the corresponding statements in the \textit{Code} model. Both models serve as input to the code analysis process described in the following section. It is important to note that this representation is not a model of the program’s runtime behavior, but rather a control flow model—that is, a structural abstraction of the code’s possible execution paths. This distinction is particularly relevant in the case of JavaScript, where asynchronous constructs may cause the actual execution order to diverge from the control structure represented in the model.

As shown in Figure~\ref{fig:graphMM}, a \textit{Control Flow} model contains a set of code subgraphs, each representing either a code block or a callable unit (e.g., a method or function). Each subgraph consists of nodes corresponding to statements, which are connected by directed edges. Every node may have outgoing and incoming edges.
Edges represent either unit calls or conditional branches. Accordingly, nodes hold references to \textit{Statement} elements in the \textit{Code} model, while edges reference either method calls or conditional expressions, also defined in the \textit{Code} model.
Since the \textit{Control Flow} model is explicitly linked to the \textit{Code} model from which it was derived, code analysis can traverse control path to access corresponding statements and perform higher-level reasoning over the program structure.

To generate \textit{Control Flow} models, we adapted the algorithm described in~\cite{aho86}. In our case, the input is a \textit{Code} model rather than an abstract syntax tree (AST), and the output is a corresponding \textit{Control Flow} model.

\begin{figure*}[!ht]
  \centering
  \includegraphics[width=0.9\textwidth]{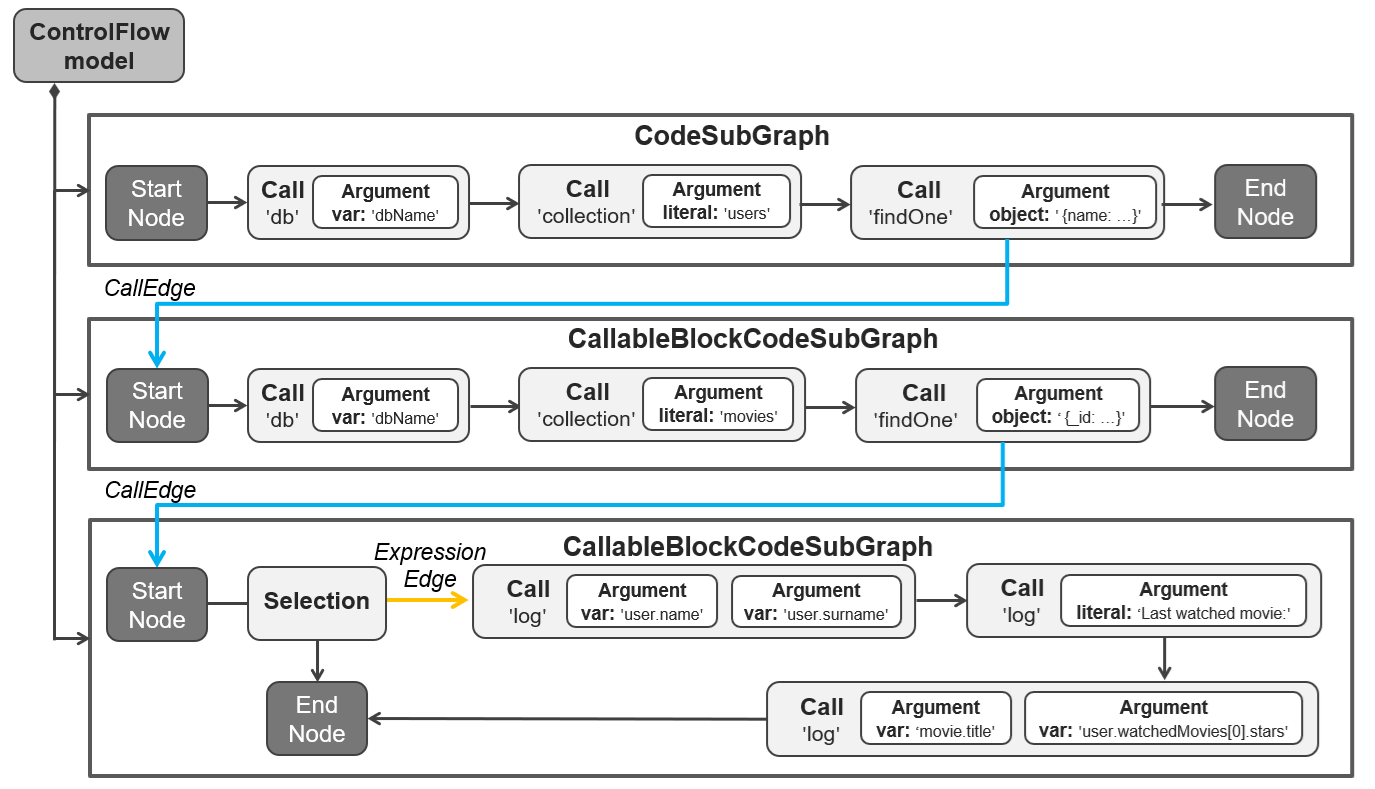}
  \caption{\textit{Control Flow} model for the the running example.}
	\label{fig:codeGraphModelRunningExample}
\end{figure*}

The \textit{Control Flow} model generated from the \textit{Code} model of the running example is shown in Figure~\ref{fig:codeGraphModelRunningExample}. The \textit{ControlFlowModel} root element aggregates three subgraphs:
The first, a \textit{CodeBlockSubGraph}, corresponds to the \textit{findOne()} method call chain
\texttt{client.db(dbName).collection ('Users').findOne({name:name}}), \textit{(nested lambda expression}).
This subgraph therefore contains three “call” nodes in addition to the \texttt{start} and \texttt{end} nodes. The third “call” node has an outgoing edge pointing to the \texttt{start} node of a \textit{CallableBlockSubGraph}, which corresponds to the \textit{findOne()} method call chain:
\texttt{client.db(dbName).collection ('Movies'). findOne({\_id:user .watched Movies[0]. movie\_id}},
\textit{(nested lambda expression}).
This second subgraph also contains three “call” nodes, with the third node connected to the \texttt{start} node of a third \textit{CallableBlockSubGraph}. This last subgraph includes a selection node, which leads to three additional “call” nodes corresponding to the three \texttt{console.log} statements within the conditional block.

\begin{figure*}[!ht]
  \centering
  \includegraphics[width=0.75\linewidth]{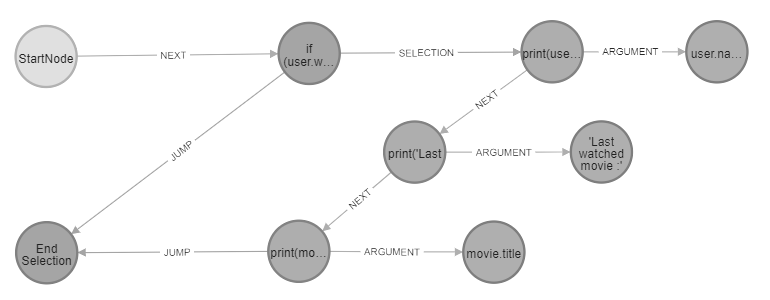}
  \caption{Graph excerpt of the Control Flow model for the running example.\label{fig:codeGraphRunningExample}}
\end{figure*}

\begin{figure*}[!ht]
  \centering
  \includegraphics[width=\textwidth]{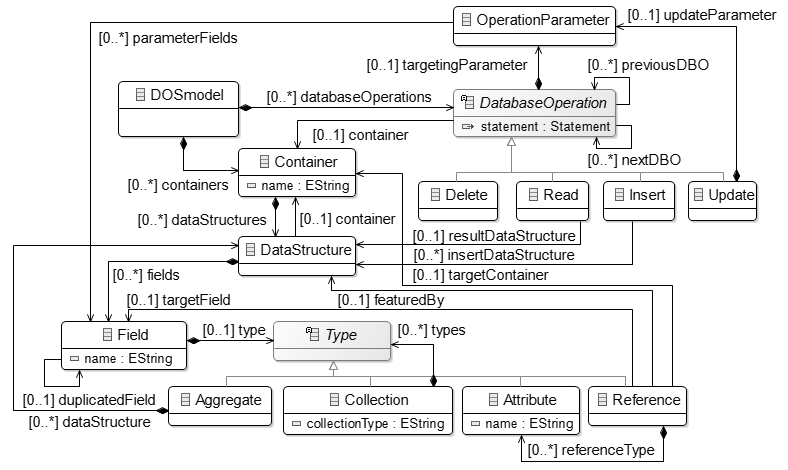}
  \caption{Database Operation\&Structure Metamodel.\label{fig:dboMetamodel}}
\end{figure*}

It is worth noting that the current algorithm does not fully handle callbacks or asynchronous control flows that are resolved only at runtime, which remain outside the current scope of our static analysis.

\paragraph{Testing}
We validated this second step by visually verifying that the generated models correctly represented the program’s control flow. For this purpose, models were stored in a Neo4J graph database and explored through the Neo4J Browser~\cite{carlos-idioms19}, which displays query results as navigable graphs. Figure~\ref{fig:codeGraphRunningExample} shows an excerpt of the Control Flow model for the running example, illustrating an if-then statement with its alternative paths and sequential \texttt{console.log} calls. These visualizations proved much more readable and intuitive than the raw model, facilitating validation. The mapping from the Control Flow model to Neo4J was straightforward, and the generation code was automatically produced using the same mechanism as for the Code model validation.

\section{Finding Operations and Structure of the database \label{sec:generatingDOS}}

\textit{Code} and \textit{Control Flow} models are analyzed to discover the implicit database schema and to apply database refactorings. In the first step of the analysis, information about CRUD operations and the structure of the manipulated data  is captured in an intermediate representation. This representation is defined by the \textit{Database Operation and Structure} (DOS) metamodel, as shown in Figure~\ref{fig:dboMetamodel}. In this section, we present the Algorithm~\ref{alg:code-analysis} designed to obtain a DOS model.

The root element of the DOS metamodel is \textit{DOSmodel}, which aggregates \textit{OperationDatabase} and \textit{Container} elements, as shown in Figure~\ref{fig:dboMetamodel}. A \textit{Container} represents a data storage unit, such as a collection in a document store or a table in a relational database. Each container holds one or more \textit{DataStructure} elements, which in turn aggregate the set of fields present in the objects stored within that container. Since NoSQL systems are often schemaless, a container may include multiple data structures to manage structural variations. However, as noted in Section~\ref{sec:introduction}, our approach does not address the detection of structural variability.

\begin{figure}[!ht]
  \centering
  \includegraphics[width=\linewidth]{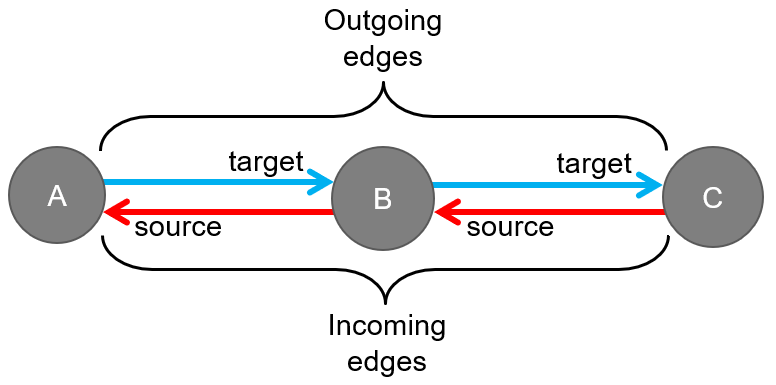}
  \caption{Nodes and edges in Control Flow models.\label{fig:graphEdges}}
\end{figure}

Each \textit{Field} has a name and a type. The type can be one of the following: \textit{Attribute}, \textit{Collection}, \textit{Aggregate}, or \textit{Reference}. An \textit{Aggregate} type encapsulates a \textit{DataStructure} that represents the internal structure of embedded objects within a root object. A \textit{Reference} type links to attributes belonging to another data structure, indicating a reference relationship between entities.
With respect to database operations, the metamodel defines a specific subclass of \textit{DatabaseOperation} for each CRUD operation: \textit{Read}, \textit{Insert}, \textit{Update}, and \textit{Delete}. Each \textit{DatabaseOperation} maintains a reference to the corresponding \textit{Statement} in the Code model. These operations also reference the data structures they interact with and may include parameters.
Furthermore, \textit{DatabaseOperation} elements are connected through the \textit{previousDatabaseOperation} and \textit{nextDatabaseOperation} relationships, forming a chain that reflects their execution order within the control flow.

A graph traversal algorithm is applied to the Control Flow model to identify all references between data elements and their involvement in database operations. As shown in Algorithm~\ref{alg:code-analysis}, a backward traversal is used to detect operations and data dependencies, while a forward traversal discovers data structures and links each operation to the data it accesses. In Control Flow models, nodes have outgoing edges pointing to target nodes, and incoming edges originating from source nodes, as defined in the Control Flow metamodel (see Figure~\ref{fig:graphMM}). The \textit{source} and \textit{target} references of edges enable the implementation of backward and forward traversals, respectively, as illustrated in Figure~\ref{fig:graphEdges}.
In this figure, a blue edge exits from node A to node B via an \textit{outgoing edge} and a \textit{target} reference, while a red edge enters node A from node B via an \textit{incoming edge} and a \textit{source} reference. 

\begin{figure*}[!ht]
  \centering
  \includegraphics[width=0.8\linewidth]{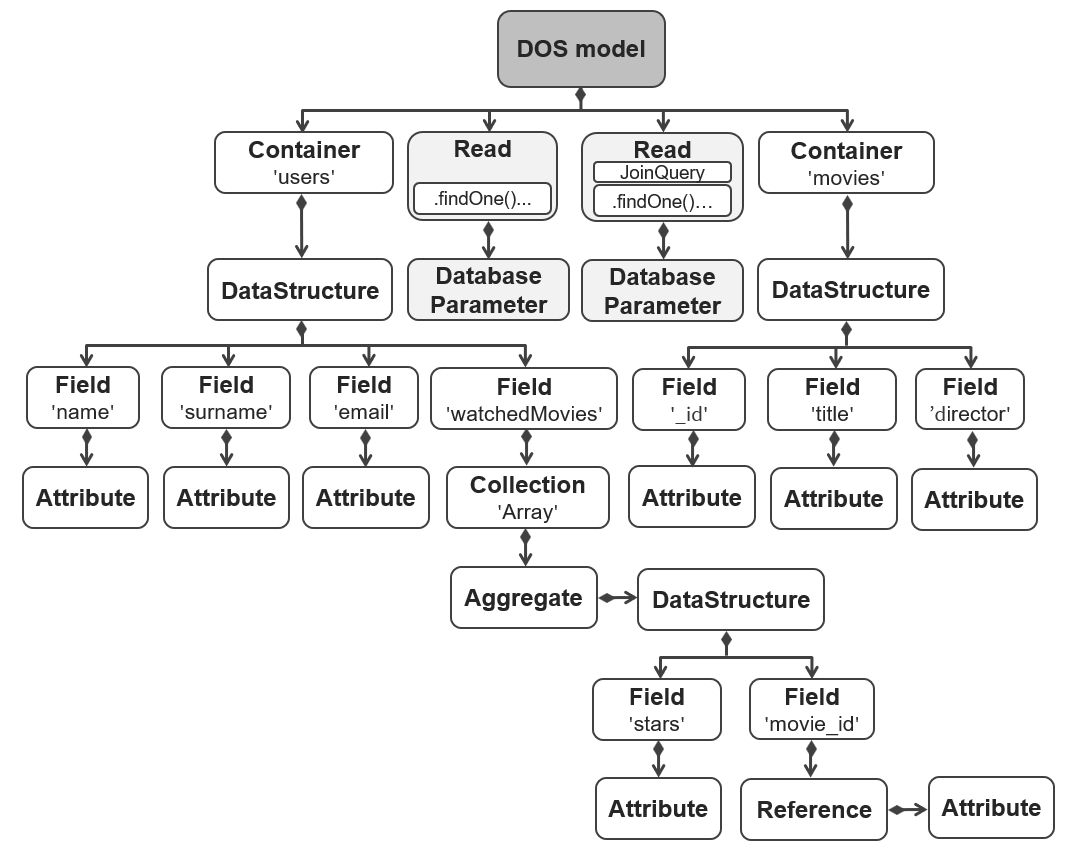}
  \caption{Database Operations \& Structure Model for the running example.\label{fig:dboModelRunningExample}}
\end{figure*}

Next, we illustrate how the DOS model is incrementally constructed through Algorithm~\ref{alg:code-analysis}, using the running example introduced in Section~\ref{sec:overview}. Table~\ref{tab:dos-algorithm} provides a compact step-by-step trace of this process: each row corresponds to a fragment of the source code, the algorithm step in which it is processed, and the resulting update to the DOS model. This stepwise mapping allows readers to visualize how the algorithm progressively shapes the final model depicted in Figure~\ref{fig:dboModelRunningExample}.

\begin{table*}[!ht]
\centering
\begin{tabular}{|c|p{3.8cm}|p{4.2cm}|p{8cm}|}
\hline
\textbf{\#} &
\textbf{Code fragment} &
\textbf{Algorithm step (line(s))} &
\textbf{DOS model update} \\
\hline
1 & \centering{-----} & Create DOS model (1) & Initialize empty DOS model. \\
\hline
2 & \texttt{findOne(users)}, \texttt{findOne(movies)} & Backward Traverse: DB Operation Creation (9). & Create two \textit{Read} object: \textit{users} and \textit{movies}, corresponding to the detected database operations. \\
\hline
3 & \texttt{\_id == user.watchedMovies[0] .movie\_id} & Backward Traverse:  Dependency Detection (18). & Detect join condition between \textit{users} and \textit{movies} based on field equality; mark \textit{movies} as a join node. \\
\hline
4 & \texttt{findOne(users)}, \texttt{findOne(movies)} & Forward Traverse: Container and DataStructure Creation (29--30). & Create two \textit{Containers}: \textit{users} and \textit{movies}, generate a corresponding DataStructure for each container, and link them to their respective \textit{Read }object. \\
\hline
5 & \texttt{user.name}, \texttt{user.surname}, \texttt{user.email} & Forward Traverse: \textit{Field} Detection (37--39). & Create fields \textit{name}, \textit{surname}, and \textit{email}, and add them to container \textit{Users}. \\
\hline
6 & \texttt{user.watchedMovies[0] .stars >= 5} & Forward Traverse: Field Detection (37--39) with new Collection and Container. & Create a \textit{Collection} and a new \textit{Container} watchedMovies, including the field \textit{stars} of type \textit{Int}. \\
\hline
7 & \texttt{movie.\_id}, \texttt{movie.title} & Forward Traverse: Field Detection (37--39). & Create \textit{id} and \textit{title} \textit{field}s in container. \\
\hline
8 & \texttt{\_id == user.watchedMovies[0] .movie\_id} & Reference Creation (48--50). & Create a reference between container \textit{users} and \textit{movies}. \\
\hline
\end{tabular}
\caption{Stepwise mapping from code fragments of the running example to algorithm steps and corresponding DOS model updates.}
\label{tab:dos-algorithm}
\end{table*}

\begin{algorithm*}[!ht]
\small
\SetAlgoLined
\LinesNumbered
\DontPrintSemicolon
\caption{Database Operation and Structure Extraction Algorithm.}\label{alg:code-analysis}
\begin{multicols}{2}

 \KwData{cfModel : Control Flow Model}
 \KwResult{dosModel : DOS Model}
$dosModel \gets createDOSmodel()$\;
$dboNodes \gets getDatabaseOpsCallNodes(cfModel)$\;
$backwardTraverse(dboNodes)$\;
$forwardTraverse(dboNodes)$\;
$createReferences(dboNodes)$\;

\;

\SetKwFunction{FMain}{backwardTraverse} \SetKwProg{Fn}{Function}{:}{}
\Fn{\FMain{$dboNodes$}}{		

	\ForEach{$dboNode \in dboNodes$}{
 		$dbo \gets createDatabaseOperation(dboNode)$\;
		$sList \gets getArguments(dboNode)$\;	
		\;
		$sNode \gets getPreviousNode(dboNode)$\;
		\While{$\exists$ $sNode$}{			
 			\uIf{$isDatabaseOperation(sNode)$ $ \land $ 
 			$getReturnVariable(sNode) \in sList$}{%
 				$pDBO \gets findDbOperation(sNode)$\;
 				$dbo.previousDBO \gets pDBO$\;
 				$pDBO.nextDBO \gets dbo$\;
 				$markJoinQuery(pDBO)$\;
  			} 
  			\ElseIf{$isAssignment(sNode)$ $ \land $  $getLeftVariable(sNode) \in sList$}{%
 				$sList.add(getRightVariable(sNode))$\;
  			}
  			
			$sNode \gets getPreviousNode(sNode)$\;
		}
	}	
}

\;

\SetKwFunction{FMain}{forwardTraverse} \SetKwProg{Fn}{Function}{:}{}
\Fn{\FMain{$dboNodes$}}{	
	$readNodes \gets getReadsOperations(dboNodes)$\;
	\ForEach{$dboNode \in readNodes$}{		
		$container \gets getOrCreateContainer()$\;
		$ds \gets getOrCreateDataStructure(container)$\;
 		$dboNode.result \gets ds$\;
		$sList \gets dboNode.statement.result$\;		
		\;	
		$tNode \gets getNextNode(dboNode)$\;
		\While{$\exists$ $tNode$}{			
 			\If{$tNode.variables \in sList$}{
 				$field \gets createField(tNode)$\;
 				$field.type \gets createType(tNode)$\;
 				$ds.fields.add(field)$
  			}
  			
			$tNode \gets getNextNode(tNode)$\;
		}
	}
}

\;

\SetKwFunction{FMain}{createReferences} \SetKwProg{Fn}{Function}{:}{}
\Fn{\FMain{$dboNodes$}}{		
	$joinQueries \gets getJoinQueries(dboNodes)$\;
	\ForEach{$dbo \in joinQueries$}{
		$sField \gets findSource(dbo.previousDBO)$\;
		$tField \gets findTarget(dbo)$\;
		$createReference(sField,tField)$\;
	}
}
\end{multicols}
\end{algorithm*}

In the following description, "line X" refers to the corresponding step in Algorithm~\ref{alg:code-analysis}, and "row Y" indicates the update of the DOS model shown in Table~\ref{tab:dos-algorithm}.

Before traversing the control-flow graph, two preliminary operations are performed. 
In an initial step, a \textit{DOSmodel} root element is created (line~1; row~1). This step is not associated with any code fragment from the running example shown in Listing~1. Next, each subgraph —representing either a function or a script— is traversed to identify the \textit{Call} nodes associated with database operations. These nodes are collected in the ordered list \texttt{dbCallNodes} (line~2). 
Relevant nodes are identified by matching function names against those defined in the data-base management API used in the code. It is important to note that in a Control Flow model, each subgraph is connected to the next in the execution sequence through a \textit{CallEdge}, as shown in Figure~\ref{fig:codeGraphModelRunningExample}. The source and target subgraphs may belong to different functions or even to different files.

After this initialization, the functions that implement the backward and forward traversals are called (lines~3~and 4). The \texttt{backwardTraverse()} function (lines~7~to~24) iterates over the \texttt{dboNodes} list, which contains the \textit{Call} nodes related to database operations.
For each visited node (\texttt{dboNode}), a \textit{Read}, \textit{Insert}, \textit{Update}, or \textit{Delete} operation is instantiated, depending on the type of database operation (line~9; row~2). In the running example, the created objects correspond to the two \texttt{findOne()} operations. The created \textit{DatabaseOperation} instance (\texttt{dbo}) contains a reference to the corresponding \textit{Statement} in the Code model, which is obtained from the node.

Next, the \textit{Argument}s of \texttt{dboNode} are stored in a variable search list \texttt{sList} (line~10), and the control flow graph is traversed backwards from the source node to the current node (lines~12-24). Each visited node (\texttt{sNode}) is processed as follows:
If the node corresponds to a database operation call and the variable receiving the result of that call matches one of the variables in \texttt{sList} (line~14), then the current operation (\texttt{dbo}) is connected to the operation created for \texttt{sNode} (\texttt{pDBO}) via the \textit{previousDBO} relationship (line~16). In turn, \texttt{pDBO} is connected back to \texttt{dbo} through the \textit{nextDBO} relationship (line~17). This bidirectional connection indicates a data dependency between the two operations—that is, the output of one serves as input to the other.
If the two operations act on different collections, the operation receiving the data is marked as a join query (line~18; row~3). In the running example, this corresponds to the equality expression between an object retrieved in the first \texttt{findOne()} operation and the filter condition of the second \texttt{findOne()}.
It is important to note that the \texttt{findDbOperation} function is responsible for retrieving the database operation call statement (\textit{Call}) referenced by the visited node (\texttt{sNode}). This function navigates from the Control Flow model to the Code model and then checks whether the call is present in the set of database operations stored in the DOS model.
If the visited node \texttt{sNode} does not meet the conditions above, and its associated \textit{Statement} is an assignment, the algorithm checks whether the variable on the left-hand side is in \texttt{sList} (line~19). If so, the variable on the right-hand side of the assignment is added to \texttt{sList} (line~20), thus continuing the tracking of data dependencies through variable propagation.

Once the backward traversal is completed, all database operation \textit{Call} nodes are revisited using a forward traversal (line~4). In this traversal, only nodes corresponding to \textit{Read} operations are processed, through the \textit{forwardTraverse()} function.
For each \textit{Read} node (line~28), a \textit{DataStructure} is instantiated and aggregated into a \textit{Container}. Both instances are created if they do not already exist (lines~29-30; row~4). In the running example, this corresponds to the creation of two containers, \textit{Users} and \textit{Movies}, each one corresponding to a distinct \texttt{findOne()} invocation. Additionally, the \textit{DataStructure} is linked to the corresponding \textit{Read} operation (line~31). A new search list is also initialized in this traversal (line~32), which is used to collect the data retrieved by the \textit{Read} operations.

Each subgraph is traversed (lines~34~to~42), starting from the node that follows the database operation node (\texttt{tNode}) (lines~34~and~35). For each visited node, its list of variables is iterated to check which variables are holding values read from the database. To do this, each variable is matched against the elements of \texttt{sList} (line~36).
When a match is found, the corresponding \textit{Read} statement is analyzed to identify the accessed fields of the retrieved object. If a property access is detected, a new \textit{Field} is created and associated with the current \textit{DataStructure} (rows~5, 6 and 7). The type of the \textit{Field} is determined as follows (line~38):

\begin{itemize}%[topsep=2pt, itemsep=1pt, parsep=0pt]
   \item If another property is accessed from the current property, this indicates that the current property holds an embedded object, and its type is set to \textit{Aggregate} (row~6). In the running example, the field \texttt{watched Movies} contains an aggregate object due to the access to its \texttt{stars} property.
   
   \item If a collection operation is found, the type of the current property is set to \textit{Collection} (row~6). In the running example, the field \texttt{watchedMovies} is recognized as a collection of aggregate objects.
   
   \item Otherwise, the type of the property is set to \textit{Attribute} (rows~5 and~7) — note that references are not detected during the forward traversal of the graph. In the running example, this applies to fields such as \texttt{name}, \texttt{surname}, and \texttt{email} in the \texttt{Users} container, and \texttt{title} in the \texttt{Movies} container. These attributes are detected because the code accesses the properties of the corresponding objects, such as \texttt{user.name} or \texttt{movie.title}.
\end{itemize}

It is important to note that the fields of a particular data structure may be discovered at any point during the traversal. As a result, the type of a field may either change or remain undetermined until sufficient context is available.

Once the forward traversal is completed, the identification of attributes that are actually references is performed by calling the \texttt{createReferences} function (line~5). This function first collects all \texttt{Read} operations marked as join queries (line~46), that is, those for which the \textit{previousDatabaseOperation} relationship is not null.
This collection is then iterated (lines~47–51), and for each join query, the join condition is analyzed to extract the name of the field involved in the join (lines~48~and~49). A \textit{Reference} type is then created in the DOS model (row~8). In the running example, this corresponds to the reference between the \texttt{Users} and \texttt{Movies} containers, established through the join condition. This reference points to the target container and is also associated with the corresponding attribute field previously identified during the forward traversal.
Since the corresponding data structures may not yet exist during the initial identification of join queries, reference relationships cannot be extracted at that stage.
Finally, the algorithm checks whether multiple data structures contain identical sets of fields. When duplicates are found, only one structure is retained and the others are discarded. Figure~\ref{fig:dboModelRunningExample} shows the DOS model obtained by applying Algorithm~\ref{alg:code-analysis} to the running example. 

\paragraph{Testing}
Algorithm~\ref{alg:code-analysis} was tested on small code snippets structurally similar to the running example. For each snippet, we manually verified that the resulting DOS model correctly captured the expected \textit{Container}, \textit{DataStructure}, \textit{Field}, and \textit{Type} elements. Starting with a minimal example containing a single Read operation, we progressively added statements manipulating the query result to uncover new fields and containers.
After each iteration, the model was rechecked to ensure that the structural changes were correctly reflected.

\section{Obtaining the Database Schema  \label{sec:obtainingschemas}}

\begin{figure*}[!htb]\centering
  \includegraphics[width=\textwidth]{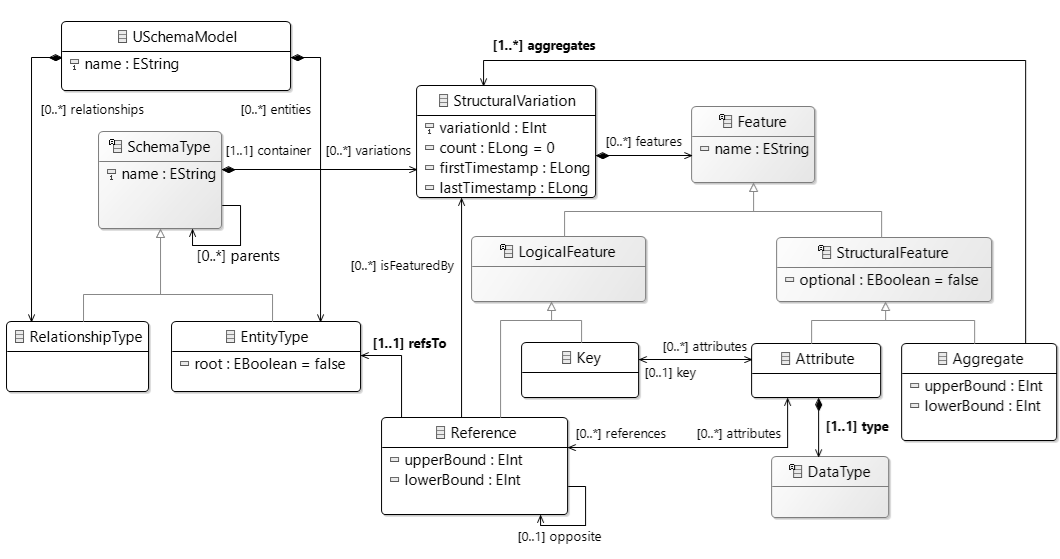}
  \caption{\uschema{} Data Model.\label{fig:uschemametamodel}}
\end{figure*}

As explained in Section \ref{sec:overview}, the DOS metamodel is designed to represent both the set of database operations found in the analyzed code and the structure of the data stored. The structural part of a DOS model captures the database’s physical schema: containers of objects whose data structure consists of a set of fields, which can be attributes, collections, aggregates, or references. From the physical schema, a logical schema can be derived. In our approach, the \uschema{} unified metamodel~\citep{metamodel2021} is used to represent the logical schema, which is obtained through a model-to-model transformation. The \uschema{} metamodel is depicted in Figure~\ref{fig:uschemametamodel}.

In \uschema{}, a schema consists of a set of schema types, which can be either entity types or relationship types. The latter include, for example, many-to-many tables in relational schemas and edges in graph schemas. Each type aggregates one or more structural variations, each composed of a set of features of two kinds: structural and logical. Structural features can be attributes or aggregates, and define the internal structure of database objects. Logical features can be keys or references, and are used to identify attributes that hold identifier values. A more detailed explanation can be found in~\cite{metamodel2021}.

\begin{table}[!ht]
\begin{center}
\begin{tabular}{lp{.65\linewidth}}
\toprule
 \textbf{DOS Model} 	&  \textbf{\uschema{}}\\
\midrule
 Container 			 	& Entity Type\\

 DataStructure 			& Structural Variation of an entity type\\

 Attribute field			& Attribute feature\\

 Reference field  		& Reference feature\\

 Aggregate field    		& (Non-root) Entity Type, Structural Variation, and a Aggregate feature\\
\bottomrule
\end{tabular}
\end{center}
\caption{DOS metamodel to \uschema{} metamodel Mappings.\label{table:dbo-to-uschema-mappings}}
\end{table}

The transformation begins by creating a \textit{USchema-Model} as the root element. Then, an \textit{EntityType} is created for each \textit{Container} element. Next, each \textit{DataStructure} is mapped to a \textit{StructuralVariation} of the \uschema{} metamodel, which includes a set of features. An entity type may have multiple variations if the code analysis is performed on different versions of the same script or application.
\textit{Field} elements are mapped to \uschema{} features as follows:
(i) An \textit{Attribute} feature is created for each \textit{Attribute} field, preserving its name and primitive type.
(ii) A \textit{Reference} feature is created for each \textit{Reference} field, which is linked to the \textit{EntityType} corresponding to the \textit{Container} specified by the \textit{targetContainer} property.
(iii) Each \textit{Composition} field is mapped to an embedded (non-root) entity type and an associated \textit{Aggregate} feature. This process also creates a \textit{StructuralVariation} for the referenced \textit{DataStructure}, and is recursively applied to all fields contained within it.
(iv) Each \textit{Collection} field is mapped to an \textit{Attribute} feature. The collection type is obtained from the \textit{collectionType} property of the \textit{Collection} meta-class. Collections may contain either primitive values or embedded objects. 

\begin{figure}[!ht]
  \centering
  \includegraphics[width=0.8\linewidth]{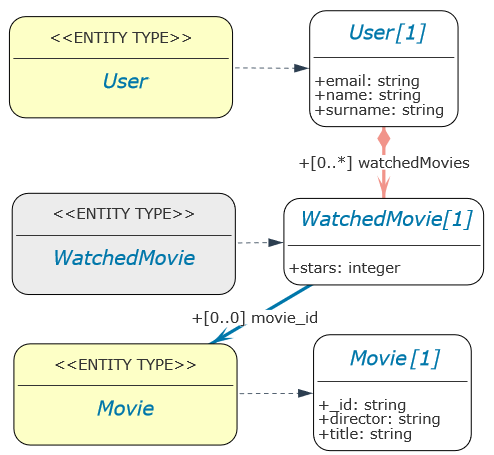}
  \caption{\uschema{} model obtained from code analysis for the running example.\label{fig:uschemaCodeAnalysisRunningExample}}
\end{figure}

Table~\ref{table:dbo-to-uschema-mappings} shows the DOS to \uschema{} mappings applied to obtain logical schemas. 
Figure~\ref{fig:uschemaCodeAnalysisRunningExample} shows the \uschema{} model derived from the DOS model presented in Figure~\ref{fig:dboModelRunningExample}. The schema is visualized using SkiQL, a notation specifically designed for NoSQL schemas~\citep{skiql2022}. 

\paragraph{Testing}
As explained in the previous sections, each transformation step was validated independently. Manual techniques were applied for the Code and DOS models, using incremental testing with code snippets of increasing complexity, and for the Control Flow models, through visual inspection of the generated graphs. In contrast, the final transformation (from DOS to U-Schema) was automatically validated using Athena~\cite{alberto-comonos2021}, a generic schema definition language built on top of U-Schema. Athena enabled automated testing by generating and comparing reference scripts, providing an objective mechanism to verify the correctness of the transformation and, consequently, of the entire workflow.

\section{Finding \textit{join removal} refactoring candidates\label{sec:generatingplans}}

A DOS model provides valuable insights that support database practitioners in identifying potential refactorings aimed at improving data quality or query performance. For example, analyzing the number of fields in each entity type can help detect overloaded entities that might benefit from being decomposed into smaller, more cohesive structures. In addition, the presence of join queries may reveal opportunities to group related data into aggregates, thereby avoiding time-consuming joins between separate containers and improving query performance.

To illustrate how our approach enables the detection and application of refactorings, we focus on the \textit{join query removal} refactoring, previously introduced in Section~\ref{sec:overview}. Applying this refactoring requires the following elements: the join queries, the two involved entity types, and the set of fields to be duplicated. Notably, this last element is the only one not directly represented in the DOS model. The remainder of this section describes how such information is identified. We refer to the data provided to practitioners to support this refactoring as a \textit{join removal plan}.

Algorithm~\ref{alg:code-analysis} identifies and marks the \textit{Read} operations that correspond to join queries (line~12), as described in Section~\ref{sec:generatingDOS}. Therefore, the main objective of the next analysis step is to determine which fields should be duplicated for each join query present in the DOS model. These fields are discovered by analyzing the statements that appear after the join query in the control flow. In particular, the algorithm looks for variable access statements in which the result variable of the join query is accessed via dot notation to retrieve specific fields from the data structure associated with the referenced container.
 It is important to note that the result variables of both queries involved in a join typically appear within the same \textit{CodeBlock}, facilitating this type of data dependency analysis.

Algorithm~\ref{alg:duplicate-finder} identifies and selects the data to be duplicated in order to eliminate a join query. It proceeds as follows. First, the set of join queries (\textit{Read} operations) is collected—specifically, those with at least one \textit{prevDatabaseOp} relationship (line~1). Next, two variable search lists are initialized (lines~2~and~3). For each join query, the result variable is added to \texttt{joinSList} (line~7), while the result variable of the preceding query is added to \texttt{prevSList} (line~8).

\begin{algorithm}[!ht]
\small
\SetAlgoLined
\LinesNumbered
\DontPrintSemicolon
\caption{Field Duplication Detection Algorithm.}\label{alg:duplicate-finder}

 \KwData{cfModel : Control Flow Model}
 \KwData{dosModel : DOS Model}
 \KwResult{dosSModel : DOS Model}
$joins \gets getJoinQueries(dosModel)$\;

$joinSList \gets \emptyset$\;
$prevSList \gets \emptyset$\;	

\;
\ForEach{$join \in joins$}{
	$pDBO \gets dbo.prevDBO$\;
	$joinSList.add(getResultVariable(join))$\;
	$prevSList.add(getResultVariable(pDBO))$\;		
	\;		
	$node \gets findFollowingNode(cfModel, join)$\;
	\While{$\exists$ $node$}{		
		\If{$node.variables \in joinSList$ $\land$ 
			 $node.variables \in prevSList$}{
			
  			$fields \gets getFields(dbo, node.variables)$\;
 			$joinQueryFields \gets copyFields(fields)$\;
 			$pDBO.resultDS.add(joinQueryFields)$\;
  		}
  		\If{$isAssignment(node, joinSList)$}{
 			$joinSList.add(node.variables)$\;
  		}
  		\If{$isAssignment(node, prevSList)$}{
 			$prevSList.add(node.variables)$\;
  		}
  			
		$node \gets getFollowingNode(node)$\;
	}
}
\end{algorithm}

At this point, the function \textit{findFollowingNode} retrieves the node from the control flow model that corresponds to the current join query and returns its immediate successor (line~10). A forward traversal is then performed to analyze the variables used in the statements that follow in the control flow (lines~11~to~23).
Each visited node is inspected to check whether any of its variables appear in both the \texttt{joinSList} and \texttt{prevSList} search lists (line~12). This condition is satisfied when the result of the join query is used alongside  the result of the preceding query. In such cases, the fields accessed from the join query’s result variable are considered candidates for duplication if the join is to be removed.

When this condition holds, the DOS model is updated as follows. First, the fields accessed in the current statement are obtained from the corresponding node in the \textit{Control Flow} model (line~13). Then, each corresponding \textit{Field} is copied (line~14) and assigned to the result of the preceding \textit{Read} operation (line~15)—that is, to the \textit{DataStructure} of the object retrieved by the prior database operation. The copied \textit{Field} maintains a reference to the original.
During the traversal, if an assignment is detected in which the result of a database operation is assigned to another variable (line~17), the new variable is added to \texttt{joinSList} (line~18). The same logic is applied to the result of the preceding operation (lines~20–21). This process is repeated for all database operations.

When the algorithm is applied to the DOS model of the running example, the join query list includes a single \textit{Read} operation, as shown in Figure~\ref{fig:dboModelRunningExample}. This operation is processed as follows.

The result variable of the join query (\textit{movie}) is added to \texttt{joinSList}, and the result variable of the previous query (\textit{user}) is added to \texttt{prevSList}. The control flow traversal then starts from the node returned by the \textit{findFollowingNode} function, which returns the node corresponding to the \texttt{if-then} statement labeled \textit{Selection} in Figure~\ref{fig:codeGraphModelRunningExample}.

Subsequent nodes are traversed to detect the usage of variables in the search lists. Specifically, the algorithm checks whether the variables \texttt{user} (in \texttt{prevSList}) and \texttt{movie} (in \texttt{joinSList}) are used together within the same \textit{Statement} or \textit{CodeBlock}. It visits the three \textit{Call} nodes representing the \texttt{console.log} statements in the control flow model and finds that both variables are used together in the last call. At this point, the \texttt{title} field is identified as a candidate for duplication, as the expression \texttt{movie.title} is detected.
As a result, the \texttt{title} field is copied and added to the \textit{DataStructure} that contains the \texttt{movie\_id} field involved in the join condition. The newly created field is marked as duplicated and maintains a reference to the original \textit{Field} in \textit{Movie}.

Once the fields to be duplicated are identified and the DOS model is updated, the model contains all the information required for a database practitioner to decide whether a particular join query should be removed. A dedicated application could be developed to display this information in the form of \textit{join removal plans}, allowing users to 
select which refactorings are automatically applied.
In the case of a join removal refactoring, each plan would include the following: the join query and related operations, the source and target entity types, the fields from the target to be duplicated in the source, and both the original and a rewritten version of the affected code. 

When a user selects a database refactoring, a schema change must be performed.
This requires updating the logical schema, the database contents, and the application code. We have implemented the automatic update of both the database and the code, as described in the following sections.

\paragraph{Updating database}

We have used the Orion language~\citep{alberto-er2021} to update both the schema and the stored data. Orion is a generic schema evolution language for NoSQL and relational databases, defined for the \uschema{} metamodel.
From each refactoring plan, an Orion operation is automatically generated. For example, a \texttt{COPY} operation could be used for the field duplication required by a join removal refactoring. This operation copies one or more fields from a source entity type to a target entity type. In the case of our running example, it would be expressed as:
\texttt{COPY  Movies::\{title,director\}} TO \texttt{Users::watchedMovies. movie\_id WHERE movie\_id = \_id}.
Each Orion operation updates the \uschema{} model via a model-to-model transformation, and code updating stored data is generated via a model-to-text transformation, which is specific for each database platform. By using Orion, we ensure that schema refactorings are applied consistently across both the schema and data layers.

\paragraph{Updating code}

Code is updated through a two-step process. First, a model-to-model transformation is applied to update the \textit{Code} model. This automatic transformation takes both the DOS and Code models as input and produces a modified \textit{Code} model as output. The DOS model provides the necessary information: the data to be duplicated for each join query and the \textit{Read} operations that can be removed. Accordingly, the identified join queries are eliminated, and the corresponding code expressions where the result variable was used are replaced.
Specifically, the \textit{CodeBlock} from the second query is moved into the first one. 
Then, all occurrences of the original result variable are replaced by the result variable of the preceding query, followed by access to the newly duplicated field. If the duplicated field belongs to an aggregate, the replacement must include access through the corresponding \textit{Aggregate} field, as illustrated in the running example.

Finally, in the second step, the updated \textit{Code} model is traversed to regenerate the source code. Note that the regenerated code is automatically modified and may contain minor issues that require manual review.

In the running example, the second query is identified as a join query, as indicated in the DOS model shown in Figure~\ref{fig:dboModelRunningExample}. This query uses the \texttt{user} result variable produced by the preceding query, as reflected in the \textit{Code} model in Figure~\ref{fig:schemaRunningExample}. If the join query is removed, the expression \texttt{movie.title} is replaced with \texttt{user.watchedMovies [0].movie\_title}, according to the duplication logic previously described.

\section{Validation}\label{sec:validation}

A testing strategy has been applied to each step of the reverse engineering process described in the previous sections. In this section, we present the validation of the complete code analysis process. The input is JavaScript code that manipulates a MongoDB database, and the output is the inferred database schema along with a list of join removal plans.
We have also considered the application of schema change operations for join removals that were assumed to be selected by practitioners, in order to simulate a realistic refactoring scenario.

First, we describe the experimental design and the me-thodology followed for the evaluation. Then, we present the results, and finally, we discuss the limitations of our validation.

\subsection{Experimental setup}

Because no publicly available datasets provide both the source code and its corresponding database schema as a reliable ground truth, we generated our own validation environment. In practice, the intended logical schema of real applications is seldom documented or formally specified, which prevents an objective assessment of the accuracy of the inferred models.

To archieve this, we generated all components of the test environment. The database schema (\uschema{} model) was created using the \textit{Athena} language~\citep{alberto-comonos2021}; the dataset was generated with the data generator \textit{Deimos}~\citep{alberto-comonos2020}. Both Athena and Deimos are tools specifically designed for working with \uschema{} models. 
The application code was automatically generated using a large language model (LLM), which was provided with the schema definition. The resulting code simulated the backend logic of a small music streaming service, following common JavaScript development statements. This strategy allowed us to create diverse and reproducible examples where the intended schema was known in advance, thus enabling a systematic validation of our approach despite the scarcity of real-world datasets. The generated project and the complete implementation of the proposed approach are publicly available in the GitHub repository referenced in Section~\ref{sec:introduction}.

\begin{figure*}[!htb]
  \centering
  \includegraphics[width=0.7\linewidth]{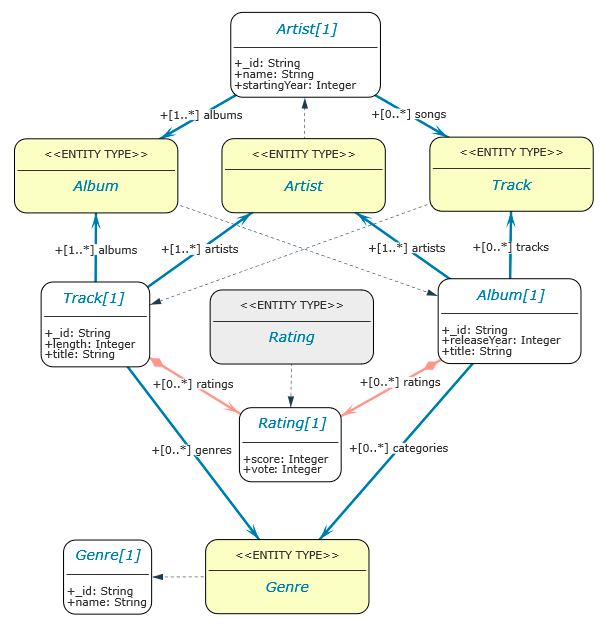}
  \caption{Validation designed schema.\label{fig:codeSchemaValidation}}
\end{figure*}

The schema designed for validation, shown in Figure~\ref{fig:codeSchemaValidation}, defines five entity types: \textit{Album}, \textit{Track}, \textit{Artist}, \textit{Rating}, and \textit{Genre}.
\textit{Artist} references zero or more \textit{Album}s and \textit{Track}s. \textit{Album} references one or more \textit{Track}s—this reference is named \texttt{songs} instead of \texttt{tracks} to demonstrate that our approach does not rely on name-based heuristics and can still accurately detect references. Similarly, both \textit{Album} and \textit{Track} reference zero or more \textit{Genre}s; however, the reference is named \texttt{categories} in \textit{Album} and \texttt{genres} in \textit{Track}.
Additionally, \textit{Album} and \textit{Track} aggregate zero or one \textit{Rating}. Thus, \textit{Rating} is an embedded entity, while the rest are root entities. As previously noted, the detection of \textit{structural variation}s through code analysis is inherently complex.
Therefore, we defined only one \textit{structural variation} per \textit{entity type}. The attributes of each entity type can be seen in Figure~\ref{fig:codeSchemaValidation}. The schema is presented using the SkiQL notation~\citep{skiql2022} as a \uschema{} model.

Based on the schema shown in Figure~\ref{fig:codeSchemaValidation}, we generated a Node.js-based backend with the assistance of ChatGPT-4o, an LLM developed by OpenAI. The model was provided with the schema as input (in the form of Figure~\ref{fig:codeSchemaValidation}) and was instructed to generate a fully functional application using the MongoDB native API driver in JavaScript, while choosing the implementation details it deemed most appropriate. Before code generation, we asked the model clarifying questions to verify that it had correctly understood all entities, attributes, and references, including their names and relationships. The resulting backend includes a REST API built with Express.js, with data stored in MongoDB and accessed via the MongoDB native API\footnote{MongoDB API: \url{https://docs.mongodb.com/drivers/node}}. The generated code included CRUD operations for all entities defined in the schema and corresponding endpoint handlers. We manually verified that all schema elements were correctly represented and handled in the generated code.

Each endpoint of the REST API performs at least one database operation. The domain model defines a class for each entity type in the schema and incorporates various structural relationships, including one-to-many and many-to-many references, as well as embedded documents. The codebase features realistic query patterns involving nested dereferencing and data aggregation, which were used to validate schema inference and the identification and application of refactorings such as join removal. In particular, several methods simulate join-like behavior by executing multiple dependent queries to resolve references across collections.

To populate the database, we created the collections \textit{artist}, \textit{album}, \textit{track}, and \textit{genre} to store the corresponding documents. Using our synthetic dataset generator, \textit{Deimos}, we generated a representative and sufficiently diverse set of instances per entity type: 10 artists, 20 albums, 60 tracks, 8 genres, and 60 ratings. This configuration ensures a representative level of variability across relationships, such as tracks with multiple genres, albums and tracks with or without ratings, and nested access patterns through artist and album references. Although the dataset is limited in size, this does not affect the validity or generalizability of our results. Our approach is based on static code analysis, and therefore operates independently of the volume of data stored in the database. What matters for schema extraction and refactoring detection is the structure and usage of the application code—not the number of records. Once the relevant code access patterns are present, the size of the dataset has no influence on the models generated.

\subsection{Methodology: Round Trip experiment}

The validation was carried out through a round-trip experiment. We used the application code generated from the previously defined database schema as input to our code analysis approach. The inferred \uschema{} model resulting from our approach was then compared with the original schema and with the schema inferred from the dataset using our data-based approach presented in~\cite{metamodel2021}. These comparisons allowed us to assess the accuracy of our solution and evaluate how it performs relative to our data-based approach.

Regarding the join removal refactoring, we manually verified that all join queries present in the application code were correctly detected. Each selected join removal was then applied, and we ensured that the schema, data, and source code were properly updated. To verify the correctness of the data updates, we checked that the duplicated fields were correctly inserted into the appropriate documents.

To validate schema updates, we modified the original schema by adding the expected duplicated fields. Specifically, for each entity type $t_1$ referencing another entity $t_2$, we added to $t_1$ the fields from $t_2$ that were selected for duplication. We then compared this manually updated schema with the one obtained after applying the automated refactoring.

To quantify the scope of the analysis, we examined all route handlers and application logic
functions included in the backend. We identified a total of 28 database-access methods distributed across five source files, covering both REST API endpoints and internal application logic. 
The backend implements full CRUD support for four entity types—Artist, Album, Track, and Genre—using the native MongoDB API. Additionally, the insert and update methods include minimal input validation, such as checks for required fields and basic type constraints. These validations reflect common practices in NoSQL applications, where the database does not enforce schema constraints, and data integrity must be ensured at the application level through presence checks and type validation. The codebase also includes six join query candidates, corresponding to references in the domain schema. These cases were manually reviewed and used to validate the ability of our solution to detect join-like access patterns and generate appropriate refactoring plans for each of them.

\subsection{Results}

Our solution successfully detected the 28 database operations present in the backend code, including those performed using the \textit{aggregate} operator. The database schema was largely inferred correctly and aligned with the predefined schema used for validation, successfully identifying all entity types, attributes, references, and aggregations. While the overall structure was accurately identified, certain data types could not be determined due to JavaScript’s dynamic typing, where type information is not always explicitly available in the code. Furthermore, all references and aggregations were inferred with a lower cardinality of 0. Determining lower-bound cardinalities from code is not feasible, as this information is often not present and is rarely enforced explicitly in real-world application logic.
As expected, the schema inferred from code analysis closely matched —but did not exactly match —the one obtained through data analysis~\citep{metamodel2021}. While data-driven approaches can detect all data types, infer lower-bound cardinalities, and identify structural variability, their heuristic-based techniques for detecting references do not guarantee full accuracy. In contrast, our code analysis approach was able to detect references precisely, overcoming the limitations of purely heuristic inference. For example, as mentioned earlier, we deliberately renamed the references from \texttt{artists} to \texttt{tracks} as \texttt{songs} and from \texttt{album} to \texttt{Genre} as \texttt{categories}, in order to expose the limitations of name-based heuristics. This reference was correctly detected by our code analysis, but was missed by the data-based approach.

Seven database operations were detected as join query candidates. Three of them follow a sequential (nested) pattern, in which the result of one query is used in a subsequent query to simulate a join. The remaining four use MongoDB's \textit{aggregate} operator to perform joins directly within the database engine. As a result, a total of eight individual queries were involved in the execution of the seven detected joins. For each detected join, our approach successfully generated a join removal plan, correctly identifying the fields to be duplicated.

\begin{table*}[ht]
\centering
\begin{tabular}{|p{0.3cm}|p{5.5cm}|l|l|p{2.2cm}|l|p{2cm}|}
\hline
\textbf{\#} & 
\makecell{\textbf{Query} \textbf{Description}} &
\makecell{\textbf{Target} \\ \textbf{Entity}} &
\makecell{\textbf{Source} \\ \textbf{Entity}} &
\makecell{\textbf{Field to} \\ \textbf{Duplicate}} &
\makecell{\textbf{Duplication} \\ \textbf{Location}} &
\makecell{\textbf{Join Type}} \\
\hline
1 & List a specific artist's album.       & \texttt{Artist}  & \texttt{Album}  & \texttt{title}                 & In \texttt{Artist}      &     Sequential Query  \\
\cline{1-7}
2 & List a specific artist's track.                 & \texttt{Artist}  & \texttt{Track}  & \texttt{title}                  & In \texttt{Artist}      & Sequential Query       \\
\cline{1-7}
3 & List all albums with its artist's name.              & \texttt{Album}  & \texttt{Artist} & \texttt{name}                  & In \texttt{Album}      & Aggregation \\
\cline{1-7}
4 & List a specific album and its tracks.       & \texttt{Album}  & \texttt{Track}  & \texttt{title}                 & In \texttt{Album}      &     Sequential Query  \\
\cline{1-7}

5 & List all albums and its genres.                 & \texttt{Album}  & \texttt{Genre}  & \texttt{name}                  & In \texttt{Album}      & Aggregation       \\
\cline{1-7}
6 & List all tracks with its album and artist.\textsuperscript{*}              & \texttt{Track}  & \texttt{Album}  & \texttt{title}, \texttt{releaseYear}        & In \texttt{Track}      & Aggregation \\
\cline{3-7}
 &               & \texttt{Track}  & \texttt{Artist} & \texttt{name}                  & In \texttt{Track}      &  Aggregation\\
\cline{1-7}
7 & List all tracks and its genres.                 & \texttt{Track}  & \texttt{Genre}  & \texttt{name}                  & In \texttt{Track}      & Aggregation       \\

\hline
\end{tabular}

\centering
\caption*{\textsuperscript{*}Query~6 includes two join operations in the same query, resulting in two duplication plans.}
\caption{Detected join removal plans based on code analysis.}
\label{tab:duplicationPlans}
\end{table*}

Table~\ref{tab:duplicationPlans} summarizes the join removal plans identified through our code analysis. Each row describes a join query detected in the codebase, detailing the source and target entity types of the join query, the fields proposed for duplication, where the duplication would be applied, and the type of join mechanism originally employed, whether through a sequential (nested) query or MongoDB’s aggregation pipeline. For instance, Query~1 retrieves a specific artist by name and then fetches all their albums using the retrieved artist information. The duplication plan suggests copying the album titles directly into the artist entity as an array to eliminate the need for repeated join operations when this information is required.

Concretely, the analysis proposed duplicating the \texttt{title} field from \texttt{Album} into \texttt{Artist} (Query 1), and from \texttt{Track} into \texttt{Artist} (Query 2), allowing each artist document to include the titles of their associated albums and tracks. The \texttt{name} attribute of the \texttt{Artist} entity was also proposed for duplication into \texttt{Album} (Query 3). Additionally, the \texttt{title} of \texttt{Track} was suggested for duplication into \texttt{Album} (Query 4), and the \texttt{name} attribute of \texttt{Genre} into both \texttt{Album} and \texttt{Track} (Queries 5 and 7). Notably, Query~6 involves two join operations—one between \texttt{Track} and \texttt{Album}, and another between \texttt{Track} and \texttt{Artist}. As a result, two duplication plans are proposed: one to copy \texttt{title} and \texttt{releaseYear} from \texttt{Album} into \texttt{Track}, and another to duplicate \texttt{name} from \texttt{Artist} into \texttt{Track}.
In such cases, when multiple fields from the same referenced entity are duplicated, they are grouped into an embedded object to preserve the semantics of the original data structure.
In such cases, when multiple fields of a referenced entity are proposed for duplication, they are grouped into an embedded object to preserve the semantic integrity of the original structure.

\subsection{Limitations of the validation}

The schema used in our validation is not overly complex; however, it was intentionally designed to include all core elements of the \uschema{} metamodel, as well as common modeling constructs and practices typically found in real-world schemas. Certain aspects were excluded from this study, such as self-references, while others—like nested queries—were included with limitations (e.g., restricted to two levels of nesting).

It is also worth noting that increasing the number of entity types or operations would not necessarily enhance the robustness of the validation. Doing so would mainly replicate the same evaluation logic across more collections, without introducing fundamentally new challenges. Although the number of schema changes required to perform join removal refactorings was relatively small, they were sufficient to demonstrate that the algorithm can accurately identify and apply field duplication when needed.

It is important to note that the projects generated with the help of LLMs are more regular and simplified than typical JavaScript applications. Real-world projects often include heterogeneous coding practices, extensive use of asynchronous constructs, and callbacks where database objects are passed across different functions or modules. Such cases are challenging for static analysis, as the control flow may only be resolved at runtime, and are not fully represented in our current evaluation dataset.  Nevertheless, the LLM-assisted experiment proved useful to identify the limits of our approach when contrasted with real-world development practices. It helped us better characterize the kinds of constructs that hinder static analysis—such as asynchronous callbacks, middleware chaining, dynamic typing, and indirect persistence layers that obscure database interactions. This limitation is explicitly acknowledged, and we outline in Section~\ref{sec:relatedWork} potential directions to mitigate it, including hybrid static/dynamic analysis and the analysis of applications written in TypeScript.

\section{Related Work}\label{sec:relatedWork}

Most of the research work conducted to date —as well as the available NoSQL tooling— has addressed the problem of schema inference by analyzing stored data, while the analysis of application code has received comparatively little attention. In this section, we contrast the static code analysis strategy proposed in this paper with the most relevant approaches to schema extraction based on both data and code analysis. We also discuss the rationale behind defining a new metamodel to represent object-oriented code, despite the existence of other metamodels for this purpose. We begin by reviewing existing proposals, grouped into three categories: data-driven strategies, code analysis approaches, and metamodels.

%%%%%%%%%%
%%%%%%%%%%
%%%%%%%%%%
\paragraph{Data-driven Schema Inference} 

A framework for schema discovery in document stores is presented in~\cite{wang-schema2015}. Based on document parsing, the approach infers a schema represented as a tree structure, capturing entities and their structural variations. The framework also includes a simple query language and a visualization mechanism to display all variations of an entity in a simplified format. The approach was validated by using several real datasets.

Another approach focused on document stores is described in~\cite{klettke-schema2015}, where the authors propose an algorithm to extract structural information from JSON data. Instead of analyzing the entire collection, a representative subset of documents is retrieved from the database. A graph is incrementally built to capture the structural features of these documents. Once the process is complete, the graph represents the union of all identified structural variations, from which a JSON Schema is generated.

Schema inference from graph stores is addressed in~\cite{comyn-wattiau2017}, where a model-driven reverse engineering approach is applied to analyze \texttt{CREATE} statements written in Neo4j's \textit{Cy-pher} query language. This analysis enables the extraction of node types, relationships, and their properties, which are then represented in a graph-based metamodel. The resulting model is subsequently transformed into an Extended Entity-Relationship (EER) schema for conceptual modeling purposes.

A proposal for schema extraction in columnar stores —specifically HBase—is presented in~\cite{frozzaDM21}, where JSON Schema is used to represent the inferred schema. The main challenge addressed is the inference of data types from byte arrays, which are the default storage format in HBase. Their method recursively analyzes the database content and applies a set of inference rules to identify data types. To validate the approach, the authors developed a publicly available tool prototype called HBase Schema Inference (HBaSI).

U-Schema was proposed by~\cite{metamodel2021} as a unified metamodel capable of representing logical schemas for both relational and NoSQL databases, including document, key–value, columnar, and graph models. The authors defined canonical forward and reverse mappings between U-Schema and each supported data model. A common strategy was established for implementing and validating the schema extraction process across all five types of databases. For validation, synthetic data were generated to populate the databases, and a four-step round-trip experiment was conducted. This USchema-based approach offers signficant advantages over the approaches discussed above that include its independence from any specific data model and its ability to extract schemas that capture both reference and aggregation relationships, as well as structural variations. Moreover, the resulting schema is a model that conforms to an Ecore/EMF metamodel~\citep{steinberg-emf2009}, which enables the use of EMF tooling to build database utilities and facilitates interoperability with other tools, as discussed in~\cite{bermudez-infsys2017}.

Building upon the same U-Schema foundation, our code analysis approach complements the data-driven strategy proposed in~\cite{metamodel2021}. While analyzing application code allows reference relationships to be directly identified —without relying on heuristics, which may not always be reliable— structural variability is more effectively discovered through data analysis. In contrast, detecting schema variations through code requires access to multiple versions of the application over time. The other advantages discussed above —such as being a generic approach, extracting schemas with references and aggregation, and producing models conforming to Ecore— also apply to our code analysis solution.

%%%%%%%%%%
%%%%%%%%%%
%%%%%%%%%%
\paragraph{Code analysis of NoSQL applications}

In~\cite{meurice2017}, the authors present an approach to support the evolution of a schema-less NoSQL data store by analyzing both the application source code and its version history. The method involves locating database queries within the code and analyzing their arguments and return values in order to infer collections, fields, types, and references between data entities.  By applying this analysis across multiple versions of the application, the authors reconstruct a \textit{historical database schema} that captures all properties that have existed over time, including their types, potential renamings, and the dates of their introduction or removal. This historical schema is visualized in tabular form, using colors and icons to highlight potential data quality issues—such as inconsistencies or deprecated fields—and to warn developers about renamed properties or collections. In contrast, our approach uses metamodel-based representations of code, supports multiple programming languages, introduces novel analysis algorithms, and leverages the extracted intermediate models to suggest and apply database refactorings.

Detecting MongoDB access operations is addressed in \cite{cherryBGMNC22}, where the CodeQL language is used to declaratively query JavaScript code. The proposed approach combines structural code analysis with heuristic rules to deal with the dynamic nature of JavaScript. Their technique achieves a precision of $78\%$ in identifying database interactions and focuses on locating access points such as queries, insertions, updates, and deletions across diverse codebases. In contrast, our work focuses on identifying both database access operations and the associated data structures, with the goal of extracting the logical schema using a data model–independent strategy. Rather than relying on CodeQL, we define a custom transformation pipeline based on intermediate metamodels, which enables schema inference and additional tasks such as database refactoring.

A taxonomy of code smells specific to MongoDB interactions in Java applications is defined in~\cite{theis-mongodb-code-smells}, where CodeQL-based static analysis techniques are used to detect the patterns included in the taxonomy. The approach is implemented in a tool capable of identifying common anti-patterns in real-world projects, helping developers detect poor coding practices related to NoSQL database access.

Finally, an extension of the Orion engine is presented in~\cite{alberto-comonos2023}, aimed at supporting code co-evolution when the database schema changes. Recall that Orion is a generic schema evolution language defined on top of \uschema{}, as described in~\cite{orion-2024}. For each operation in the Orion taxonomy, a model-to-text transformation is implemented. This transformation takes as input an Orion model —obtained by injecting Orion scripts— and generates corresponding CodeQL queries to identify and modify the affected parts of the application code. Depending on the scenario, changes can be applied automatically or developer-facing suggestions can be generated. 

Although our work focuses on schema inference from code rather than code adaptation, both approaches share the use of \uschema{}, code analysis, and MDE techniques. The approach presented in this paper could serve as a front-end for extracting schema information prior to applying Orion operations, particularly in the context of schema-less NoSQL stores.

Two previous works from our group are also closely related to the present approach.
Sánchez-Ramón et al.~\cite{oscar2011eventhandlers} proposed a reverse engineering strategy to recover the event-handling logic of Oracle Forms applications. Their work analyzed PL/SQL triggers and identified code fragments corresponding to interaction idioms, building a Control Flow model based on Ullmann’s algorithm~\cite{aho86}. The resulting model represented control dependencies between UI, control, and business-logic layers. Although the goal was not database schema extraction, this work shares with ours the use of static analysis and control-flow reconstruction to reveal implicit structures in legacy code.

Fernández-Candel et al.~\cite{carlos-idioms19} presented a model-driven reengineering process for migrating PL/SQL triggers to Java organized into MVC layers. That work defined Code and Action models compliant with KDM and employed a testing strategy similar to the one used here to validate model transformations. The current proposal extends these ideas to a different domain—database access reconstruction—by defining a specialized Code metamodel and a chain of model transformations that culminate in the automated derivation of database schemas.

%%%%%%%%%%
%%%%%%%%%%
%%%%%%%%%%
\paragraph{Code Metamodels}

Several metamodels have been proposed to represent the structure and behavior of source code in a language-independent manner. These metamodels are widely used in static analysis, reverse engineering, and software modernization tasks, as they provide an abstract representation of code elements such as classes, methods, variables, and control flow. 
In the following, we briefly describe two of the most relevant metamodels —KDM and MoDisco— which have influenced the design of our Code metamodel, as explained at the beginning of Section~\ref{injectingcode}, and are widely recognized in the context of model-driven reverse engineering.

KDM is a specification defined by the Object Management Group~\citep{kdm}. It is a comprehensive metamodel composed of multiple packages that enable the representation of various aspects of software systems, ranging from source code to physical deployment. KDM is designed to support a wide variety of programming languages, and it provides support to be extended  in order to capture language-specific constructs. 
While our Code metamodel was inspired by the language-independent design of KDM’s Code package, it was intentionally kept simpler and focused on representing the core object-oriented constructs required for analyzing database-intensive applications.

MoDisco~\citep{modisco} is a model-driven reverse engineering framework developed as an open-source Eclipse project. It was designed to extract information from legacy applications to support their understanding, maintenance, and migration. Although initially focused on Java, MoDisco has been extended to support other languages such as C\#, and includes several \textit{Discoverers} to facilitate the injection of source code into models. Its comprehensive Java metamodel influenced the design of our Code metamodel, particularly in the representation of structural elements commonly found in object-oriented languages.

Table~\ref{tab:related-work} compares the approaches according to their analyzed artefacts (Input), produced artefacts (Output), analysis technique, intermediate representations, and their degree of language and database independence. The last column (Application) indicates how the inferred models are used or applied, when applicable. While data-driven approaches are inherently constrained by the coverage and representativeness of the available data—that is, they can only infer structures present in the analyzed instances—, code-based approaches face challenges with language dependence and dynamic constructs, and refactoring methods are often technology-specific. 
In contrast, our proposal uniquely combines static code analysis with dedicated metamodels and a language- and database-independent representation, enabling systematic schema inference and technology-agnostic refactoring.

\newcolumntype{L}[1]{>{\raggedright\arraybackslash}p{#1}}

\begin{table*}[!ht]
\centering
\begin{tabular}{|L{2.6cm}|p{1.7cm}|p{1.7cm}|p{1.5cm}|p{2.75cm}|p{2.5cm}|p{2cm}|}
\hline
\makecell{\textbf{Work}} & 
\makecell{\textbf{Input}} &
\makecell{\textbf{Output}} &
\textbf{Kind of analysis} &
\textbf{Intermediate representation} &
\textbf{Language/DB independence} &
\textbf{Application} \\
\hline
Wang et al. (2015)~\citep{wang-schema2015} %-- Schema Management for Document Stores
 & Database content & JSON Schema & Data-based & Tree & No/No (MongoDB) & \makecell[l]{Schema \\ inference} \\
\hline
Klettke et al. (2015)~\citep{klettke-schema2015} %-- Schema Extraction and Structural Outlier Detection for JSON-based NoSQL Data Stores
 & Database sample & JSON Schema & Data-based & Graph & No/No (MongoDB) & \makecell[l]{Schema \\ inference} \\
\hline
Comyn-Wattiau et al. (2017)~\citep{comyn-wattiau2017} %-- Model Driven Reverse Engineering of NoSQL Property Graph Databases: The Case of Neo4j
 & CREATE statements & EER schema & Data-based & Graph-based metamodel & No/No (Neo4j) & \makecell[l]{Schema \\ inference} \\
\hline
Frozza et al. (2021)~\citep{frozzaDM21} %-- An Approach for Schema Extraction of NoSQL Columnar Databases: The HBase Case Study
 & Database content & JSON Schema & Data-based & --- & No/No (HBase) & \makecell[l]{Schema \\ inference} \\
\hline
Fernández-Candel et al. (2022)~\citep{metamodel2021} %-- A Unified Metamodel for NoSQL and Relational Databases
& Database content & U-Schema model %(struct. variations detected)} 
& Data-based & Metamodels for Relational and NoSQL paradigms & Yes/Yes % (Relational and NoSQL)
& \makecell[l]{Schema \\ inference} \\
\hline
Cherry et al. (2022)~\citep{cherryBGMNC22}  %-- Static Analysis of Database Accesses in MongoDB Applications
& Source Code (JavaScript) & Access map & Static (code-driven) & --- & No/No (MongoDB) & Query analysis \\
\hline
Bernard et al. (2021)~\citep{theis-mongodb-code-smells} %-- MongoDB Code Smells: Defining, Classifying and Detecting Code Smells for MongoDB Interactions in Java Programs
 & Source Code (Java) & \makecell[l]{Code \\ patterns} & Static (code-driven) & --- & No/No (MongoDB) & Query analysis \\
\hline
Meurice et al. (2017)~\citep{meurice2017} %-- Supporting Schema Evolution in Schema-less NoSQL Data Stores
& Source Code (Java) & Schema evolution model & Static (code-driven) & --- & No/No (MongoDB) & Database evolution \\
\hline
Hernández-Chillón et al. (2023)~\citep{alberto-comonos2023} %-- Propagating Schema Changes to Code: An Approach Based on a Unified Data Model
 & Source Code (Java) & Database Changes model & Static (code-driven) & Orion metamodel & Yes/No (MongoDB) & Schema–code co-evolution \\
\hline
Sánchez Ramón et al. (2011)~\citep{oscar2011eventhandlers} %-- Reverse Engineering of Event Handlers
 & Event Handlers (PL/SQL) & Java Code & Static (code-driven) & Behaviour and Event-Handler metamodels & Yes/No (Oracle) & Software Reverse Engineering \\
\hline
Fernández-Candel Carlos et al.~\citep{carlos-idioms19} %-- Developing a model-driven reengineering approach for migrating PL/SQL triggers to Java: A practical experience
 & Source Code (PL/SQL) & Java Code & Static (code-driven) & KDM (Code and Action) metamodel & Yes/No (Oracle) & Software Reverse Engineering \\
\hline
\textbf{Our approach} & Source Code (Javascript) & U-Schema model %(no variation detection)
+ Refactoring plan & Static (code-driven) & Code, Control Flow \& DOS metamodels & Yes/Yes %(Relational and NoSQL)
& Schema inference and refactoring \\
\hline
\end{tabular}
\caption{Comparative analysis of existing approaches addressing database schema inference and source code analysis.}
\label{tab:related-work}
\end{table*}

\section{Conclusions and Future Work\label{sec:conclusions}}

This work introduced a code analysis approach to infer database schemas and generate duplication plans for data-intensive applications. Based in model-driven engineering techniques, the approach includes three key algorithms: (i) to extract the control flow, (ii) to identify database operations and the data involved, and (iii) to analyze these operations to infer the schema and propose duplication plans. Each step was evaluated individually, and the complete process was validated through a round-trip experiment.

While most existing schema inference approaches for NoSQL databases are data-driven, some recent studies have explored code-based analysis. However, these typically focus on identifying database access points rather than extracting comprehensive logical schemas. Our approach advances this direction by enabling the discovery of full schemas, including references and aggregations. This capability is particularly relevant for NoSQL systems like MongoDB, where references are often implicit—expressed in the application code but not physically stored in the database. As such, our code-driven strategy can find relationships that would remain hidden using traditional data-driven inference. Furthermore, the reference information obtained through code analysis was also used to improve database performance by identifying unnecessary joins and suggesting data duplication as an optimization strategy.

The implementation was developed for JavaScript applications using the native MongoDB API. One of the main challenges addressed was type detection in a dynamically typed language. Although the current implementation targets JavaScript and MongoDB, the proposed metamodels and algorithms are not tied to any specific language or API. The approach can be adapted to other database systems with minimal modifications. Porting to a different programming language would require a suitable parser and possibly extensions to the code metamodel; however, the core metamodel already supports the most common constructs found in object-oriented languages, facilitating cross-platform applicability.

One important limitation of our current static analysis is its handling of callbacks and asynchronous programming patterns, which are particularly frequent in JavaScript. Our algorithm can follow explicit calls in the same module, more complex situations—such as objects passed to external callbacks, asynchronous chains, or dynamically registered handlers—cannot be fully resolved statically, since the exact control flow may only be determined at runtime. 

\paragraph{Future Work} 
Applying the analysis process to different versions of a codebase would enable tracking the evolution of the database schema over time and detecting structural variability. This analysis could reveal changes in entity types, additions or removals of fields, data type modifications, and updates to relationships such as references or aggregations. It would also help detect cases where instances of the same object type (e.g., documents in the same collection) exhibit different structures due to changes introduced in the code across versions. Identifying such variations would be key to understanding the application’s evolution, evaluating the impact of refactorings, and ensuring both backward compatibility and data consistency in NoSQL systems.

The generation of duplication plans could be further enhanced by integrating more sophisticated heuristics or even by automating the selection of optimal plans. This could involve testing various duplication strategies against actual application workloads and measuring their performance impact. To preserve data consistency, the process could also include the generation of code that ensures synchronized updates to duplicated fields when the source data changes.

Recent advances in large language models (LLMs) specialized in code understanding —such as OpenAI’s Codex, CodeLlama, or Anthropic’s Claude Sonnet 4.5— open pro-mising opportunities to complement static analysis. These models can reason about implicit semantics, incomplete code fragments, and dynamically generated constructs that are difficult to capture through purely syntactic or control-flow–based methods. However, their outputs are non-deter-ministic, context-dependent, and lack traceability, which complicates reproducibility and validation—key aspects in both research and industrial contexts. Moreover, LLMs require substantial computational resources and may introduce biases or inaccuracies when reasoning over complex or domain-specific codebases.

A promising research direction is to develop hybrid approaches that integrate LLM-based reasoning with static analysis. In such a combination, static analysis would provide systematic coverage and formal guarantees, while LLMs could contribute higher-level reasoning and pattern recognition capabilities, achieving a balance between reliability and flexibility. Hybrid static/dynamic solutions, such as workload monitoring or runtime tracing, could also improve the accuracy of schema inference in realistic settings where control flow depends on runtime conditions.

Our current validation is based on synthetic projects generated with the assistance of LLMs. This strategy allowed us to control the scenarios and obtain ground truth, but the resulting code was more regular than typical Java-Script applications. Real-world projects often exhibit heterogeneous coding practices, including extensive use of asynchronous patterns and callbacks where query results are passed across functions or modules. Such cases are only partially captured by our static analysis, since the exact control flow may only be determined at runtime. Finally, future work will extend the base metamodel by adding language-specific extensions. We plan to start with JavaScript and later experiment with Python and TypeScript, which will require implementing injectors from their ASTs to the Code model. Supporting multi-paradigm languages such as Scala would be considerably more complex, as it would involve modeling functional and hybrid constructs beyond the current object-oriented scope.

\section*{Acknowledgements}
The contributions of the first and third authors were supported by project PID2020-117391GB-I00, funded by MICIU/AEI /10.13039/501100011033 (Spain), and co-fun-ded by ERDF/EU;
Anthony Cleve was supported by the Fonds de la Recherche Scientifique (F.R.S.-FNRS) under the PDR project INSTINCT (35270712).

\bibliographystyle{plainnat}
{\small \bibliography{main}}

@article{orion-2024,
  author={Chillón, Alberto Hernández and Klettke, Meike and Ruiz, Diego Sevilla and Molina, Jesús García},
  journal={IEEE Transactions on Knowledge and Data Engineering}, 
  title={A Generic Schema Evolution Approach for NoSQL and Relational Databases}, 
  year={2024},
  volume={36},
  number={7},
  pages={2774-2789},
  doi={10.1109/TKDE.2024.3362273}}

@inproceedings{alberto-er2021,
  author    = {Alberto {Hernández Chillón} and Diego {Sevilla Ruiz} and Jesús {Garcia-Molina}},
  title     = {{Towards a Taxonomy of Schema Changes for NoSQL Databases: The Orion Language}},
  booktitle = {Conceptual Modeling - {ER} 2021 40th Int. Conf. on Conceptual Modeling, St.John's, NL, Canada},
  pages     = {176-185},
  year      = {2021},
  doi       = {10.1007/978-3-030-89022-3_15},
  volume    = {13011}
}

@inproceedings{alberto-comonos2020,
  author    = {Alberto {Hern{\'{a}}ndez Chill{\'{o}}n} and Diego {Sevilla Ruiz} and Jes{\'{u}}s Garc{\'{i}}a-Molina},
  title     = {{Deimos: A Model-based NoSQL Data Generation Language}},
  booktitle = {Advances in Conceptual Modeling - {ER} 2020 Workshops CoMoNoS, Viena, Austria},
  pages     = {151-161},
  year      = {2020},
  doi       = {10.1007/978-3-030-65847-2\_14},
  volume    = {12584}
}

@book{volter-book,
  author    = {Markus Voelter and
               Sebastian Benz and
               Christian Dietrich and
               Birgit Engelmann and 
               Mats Helander and
               Lennart Kats and
               Eelco Visser and
               Guido Wachsmuth},
  title     = {{DSL} Engineering - Designing, Implementing and Using Domain-Specific
               Languages},
  publisher = {DSLbook.org},
  year      = {2013},
  url       = {http://www.dslbook.org},
  isbn      = {978-1-4812-1858-0}
}

@InProceedings{sevilla-er2015,
  author      = {Diego {Sevilla Ruiz} and Severino {Feliciano Morales} and Jes\'us {Garc\'ia Molina}},
  title       = {{Inferring Versioned Schemas from NoSQL Databases and Its Applications}},
  booktitle   = {34th International Conference on Conceptual Modeling (ER)},
  pages       = {467--480},
  month       = {October},
  year        = 2015,
  address     = {Stockholm, Sweden}
}

@Book{fowler-dsl2010,
  author      = {Martin Fowler},
  title       = {{Domain-Specific Languages}},
  publisher   = {Addison-Wesley},
  year        = 2010
}

@Book{fowler-nosql2012,
  author      = {Pramod Sadalage and Martin Fowler},
  title       = {{NoSQL Distilled. A Brief Guide to the Emerging World of Polyglot Persistence}},
  publisher   = {Addison-Wesley},
  year        = 2012
}

@Book{steinberg-emf2009,
  author      = {David Steinberg and Frank Budinsky and Marcelo Paternostro and Ed Merks},
  title       = {{EMF: Eclipse Modeling Framework 2.0}},
  year        = 2009,
  publisher   = {Addison-Wesley Professional}
}

@article{wang-schema2015,
  author    = {Lanjun Wang and
               Oktie Hassanzadeh and
               Shuo Zhang and
               Juwei Shi and
               Limei Jiao and
               Jia Zou and
               Chen Wang},
  title     = {Schema Management for Document Stores},
  journal   = {Proc. {VLDB} Endow.},
  volume    = {8},
  number    = {9},
  pages     = {922--933},
  year      = {2015},
  doi       = {10.14778/2777598.2777601},
}

@inproceedings{meurice2017,
  author    = {Loup Meurice and
               Anthony Cleve},
  title     = {Supporting schema evolution in schema-less NoSQL data stores},
  booktitle = {{IEEE} 24th International Conference on Software Analysis, Evolution
               and Reengineering, {SANER} 2017, Klagenfurt, Austria, February 20-24,
               2017},
  pages     = {457--461},
  year      = {2017}
}

@Misc{bacvanski-datav2015,
  author =       {Vladimir Bacvanski and Charles Roe},
  title =        {{Insights into NoSQL Modeling: A Dataversity Report}},
  institution =  {DataVersity Education, LLC},
  year =         2015
}

@Book{brambilla2012,
  author      = {Marco Brambilla and Jordi Cabot and Manuel Wimmer},
  title       = {{Model-Driven Software Engineering in Practice}},
  publisher   = {Morgan \& Claypool Publishers},
  year        = 2012
}

@InProceedings{klettke-schema2015,
  author      = {Meike Klettke and Uta Störl and Stefanie Scherzinger},
  title       = {{Schema Extraction and Structural Outlier Detection for JSON-based NoSQL Data Stores}},
  booktitle   = {Conference on Database Systems for Business, Technology, and Web (BTW)},
  pages       = {425--444},
  year        = 2015
}

@article{bermudez-infsys2017,
  author    = {Francisco Javier Bermudez Ruiz and
               Jes{\'{u}}s Garc{\'{\i}}a Molina and
               Oscar D{\'{\i}}az Garc{\'{\i}}a},
  title     = {On the application of model-driven engineering in data reengineering},
  journal   = {Information Systems},
  volume    = {72},
  pages     = {136--160},
  year      = {2017},
  doi       = {10.1016/j.is.2017.10.004},
}

@inproceedings{comyn-wattiau2017,
  author    = {Isabelle Comyn{-}Wattiau and Jacky Akoka},
  title     = {Model driven reverse engineering of NoSQL property graph databases: The case of Neo4j},
  booktitle = {2017 {IEEE} International Conference on Big Data, 2017, Boston, MA, USA, December 11-14, 2017},
  pages     = {453--458},
  year      = {2017}
}

@inproceedings{oscar2011eventhandlers,
  author    = {{\'{O}}scar S{\'{a}}nchez Ram{\'{o}}n and
               Jes{\'{u}}s S{\'{a}}nchez Cuadrado and
               Jes{\'{u}}s Garc{\'{\i}}a Molina},
  editor    = {Martin Pinzger and
               Denys Poshyvanyk and
               Jim Buckley},
  title     = {Reverse Engineering of Event Handlers of RAD-Based Applications},
  booktitle = {18th Working Conference on Reverse Engineering, {WCRE} 2011, Limerick,
               Ireland, October 17-20, 2011},
  pages     = {293--302},
  publisher = {{IEEE} Computer Society},
  year      = {2011},
  doi       = {10.1109/WCRE.2011.43}
}

@book{aho86,
  author    = {Alfred Aho and
               Jeffrey Ullman and
               Ravi Sethi and
               Monica Lam},
  title     = {Compilers: Principles, Techniques and Tools},
  publisher = {Addison-Wesley},
  year      = {1986},
  isbn      = {0-201-10088-6},
}

@Manual{kdm,
  title =     {{Knowledge Discovery Meta-Model (KDM)}},
  author =    {Object Management Group OMG},
  year =      2011,
  note =      {Document formal/2011-08-04.},
	url  =      {http://www.omg.org/spec/KDM/1.3}}

@article{modisco,
author = {Bruneliere, Hugo and Cabot, Jordi and Dup{\'e}, Gr{\'e}goire and Madiot, Fr{\'e}d{\'e}ric},
year = {2014},
month = {08},
pages = {1012--1032},
title = {MoDisco: a Model Driven Reverse Engineering Framework},
volume = {56},
journal = {Information and Software Technology}
}

@inproceedings{alberto-comonos2021,
  author    = {Alberto {Hernández Chillón} and Diego {Sevilla Ruiz} and Jesús Garcia-Molina},
  title     = {{Athena: A Database-Independent Schema Definition Language}},
  booktitle = {Advances in Conceptual Modeling - {ER} 2021 Workshops CoMoNoS, St.John's, NL, Canada},
  pages     = {33-42},
  year      = {2021},
  doi       = {10.1007/978-3-030-88358-4},
  volume    = {13012}
}

@article{metamodel2021,
  author    = {Carlos Javier Fern{\'{a}}ndez-Candel and
               Diego Sevilla Ruiz and
               Jes{\'{u}}s Joaqu{\'{\i}}n Garc{\'{\i}}a Molina},
  title     = {A Unified Metamodel for NoSQL and Relational Databases},
  journal = {Information Systems},
  volume = {104},
  pages = {101898},
  year = {2022},
  issn = {0306-4379},
  doi = {10.1016/j.is.2021.101898},
}

@book{greenfield-2004,
  author = {Greenfield, Jack and Short, Keith and Cook, Steve and Kent, Stuart},
  title = {Software Factories: Assembling Applications with Patterns, Models, Frameworks, and Tools},
  year = {2004},
  isbn = {0471202843},
  publisher = {John Wiley},
}

@article{carlos-idioms19,
  author    = {Carlos Javier Fern{\'{a}}ndez-Candel and
               Jes{\'{u}}s Garc{\'{\i}}a Molina and
               Francisco Javier Bermudez Ruiz and
               Jos{\'{e}} Ram{\'{o}}n Hoyos Barcel{\'{o}} and
               Diego Sevilla Ruiz and
               Benito Jos{\'{e}} Cuesta Viera},
  title     = {Developing a model-driven reengineering approach for
  			   migrating {PL/SQL} triggers to Java: {A} practical
  			   experience},
  journal   = {Journal Systems and Software},
  volume    = {151},
  pages     = {38--64},
  year      = {2019},
  doi       = {10.1016/j.jss.2019.01.068},
}

@article{skiql2022,
  author    = {Carlos Javier Fern{\'{a}}ndez-Candel and
               Diego Sevilla Ruiz and
               Jes{\'{u}}s Joaqu{\'{\i}}n Garc{\'{\i}}a-Molina},
  title     = {SkiQL: A Unified Schema Query Language},
  journal = {Data \& Knowledge Engineering},
  year      = {2022},
  volume = {148},
  doi = {10.1016/j.datak.2023.102234},
}

@MastersThesis{theis-mongodb-code-smells,
  author      = {{J}ehan Bernard and {Thomas} Kintziger},
  title       = {{TMongoDB Code Smells: Defining, Classifying and Detecting Code Smells for MongoDB Interactions in Java Programs}},
  school      = {University of Namur},
  year        = 2021,
  address     = {Namur, Belgium}
}

@inproceedings{cherryBGMNC22,
  author       = {Boris Cherry and
                  Pol Benats and
                  Maxime Gobert and
                  Loup Meurice and
                  Csaba Nagy and
                  Anthony Cleve},
  title        = {Static Analysis of Database Accesses in MongoDB Applications},
  booktitle    = {{IEEE} International Conference on Software Analysis, Evolution and
                  Reengineering, {SANER} 2022, Honolulu, HI, USA, March 15-18, 2022},
  pages        = {930--934},
  publisher    = {{IEEE}},
  year         = {2022},
  doi          = {10.1109/SANER53432.2022.00111},
}

@article{frozzaDM21,
  author       = {Angelo Augusto Frozza and
                  Eduardo Dias Defreyn and
                  Ronaldo dos Santos Mello},
  title        = {An Approach for Schema Extraction of NoSQL Columnar Databases: the
                  HBase Case Study},
  journal      = {Journal of Information and Data Management},
  volume       = {12},
  number       = {5},
  year         = {2021},
  doi          = {10.5753/jidm.2021.1966},
}

@inproceedings{alberto-comonos2023,
  author       = {Alberto Hern{\'{a}}ndez Chill{\'{o}}n and
                  Jes{\'{u}}s Garc{\'{\i}}a Molina and
                  Jos{\'{e}} Ram{\'{o}}n Hoyos and
                  Mar{\'{\i}}a{-}Jos{\'{e}} Ort{\'{\i}}n{-}Ib{\'{a}}{\~{n}}ez},
  editor       = {George Fletcher and
                  Verena Kantere},
  title        = {Propagating Schema Changes to Code: An Approach Based on a Unified
                  Data Model},
  booktitle    = {Proceedings of the Workshops of the {EDBT/ICDT} 2023 Joint Conference,
                  Ioannina, Greece, March, 28, 2023},
  series       = {{CEUR} Workshop Proceedings},
  volume       = {3379},
  publisher    = {CEUR-WS.org},
  year         = {2023},
}

\end{document}